\documentclass[11pt,english]{article}
\usepackage{rotating}
\usepackage{axodraw}
\usepackage{epsfig}
\usepackage{latexsym}
\usepackage{graphicx}
\usepackage{psfrag}
\usepackage{amsmath,amssymb}
\makeatletter
\def\section{\@startsection {section}{1}{\z@}{-3.5ex plus -1ex minus
 -.2ex}{2.3ex plus .2ex}{\large\bf}}
\def\subsection{\@startsection{subsection}{2}{\z@}{-3.25ex plus -1ex
minus -.2ex}{1.5ex plus .2ex}{\normalsize\bf}}
\makeatother
\makeatletter
\def\theequation{\arabic{section}.\arabic{equation}}

\@addtoreset{equation}{section}
\renewcommand{\theequation}{\thesection.\arabic{equation}}
\makeatother

\newcommand{\captionfonts}{\small}
\makeatletter  
\long\def\@makecaption#1#2{%
  \vskip\abovecaptionskip
  \sbox\@tempboxa{{\captionfonts #1: #2}}%
  \ifdim \wd\@tempboxa >\hsize
    {\captionfonts #1: #2\par}
  \else
    \hbox to\hsize{\hfil\box\@tempboxa\hfil}%
  \fi
  \vskip\belowcaptionskip}
\makeatother   

\catcode`@=11
\def\marginnote#1{}
\newcount\hour
\newcount\minute
\newtoks\amorpm
\hour=\time\divide\hour
by60
\minute=\time{\multiply\hour by60 \global\advance\minute
by-\hour}
\edef\standardtime{{\ifnum\hour<12 \global\amorpm={am}
\else\global\amorpm={pm}\advance\hour by-12 \fi
 \ifnum\hour=0
\hour=12 \fi
 \number\hour:\ifnum\minute<10
0\fi\number\minute\the\amorpm}}
\edef\militarytime{\number\hour:\ifnum\minute<10
0\fi\number\minute}
\def\draftlabel#1{{\@bsphack\if@filesw
{\let\thepage\relax
 \xdef\@gtempa{\write\@auxout{\string
\newlabel{#1}{{\@currentlabel}{\thepage}}}}}\@gtempa
 \if@nobreak
\ifvmode\nobreak\fi\fi\fi\@esphack}
\gdef\@eqnlabel{#1}}
\def\@eqnlabel{}
\def\@vacuum{}
\def\draftmarginnote#1{\marginpar{\raggedright\scriptsize\tt#1}}
\def\draft{\oddsidemargin
0.0truein
 \def\@oddfoot{\sl preliminary draft \hfil
\rm\thepage\hfil\sl\today\quad\militarytime}
 \let\@evenfoot\@oddfoot
\overfullrule 3pt
 \let\label=\draftlabel
\let\marginnote=\draftmarginnote
\def\@eqnnum{(\theequation)\rlap{\kern\marginparsep\tt\@eqnlabel}
\global\let\@eqnlabel\@vacuum}
}
\catcode`@=12
\def\bea{\begin{eqnarray}} \def\eea{\end{eqnarray}}
\def\be{\begin{eqnarray}} \def\ee{\end{eqnarray}} 

\newcommand{\sbe}{\sin\beta}
\newcommand{\cbe}{\cos \beta}
\newcommand{\tbe}{\tan \beta}

\newcommand{\tm}{\widetilde m}

\newcommand{\str}{{\tilde t_R}}

\def\gtiluq{{\tilde g}_u^2}
\def\gtildq{{\tilde g}_d^2}
\def\gtilupq{{\tilde g}_u^{\prime 2}}
\def\gtildpq{{\tilde g}_d^{\prime 2}}

\begin{document}

\thispagestyle{empty}

\begin{center}
\hfill UAB-FT-645 \\
\hfill FERMILAB-PUB-08-152-T \\
\hfill ANL-HEP-PR-08-32 \\
\hfill EFI-08-15
\begin{center}

\vspace{1.7cm}

{\LARGE\bf The Effective Theory of the Light Stop Scenario}

\end{center}

\vspace{1.4cm}

{\Large \bf M. Carena$^{\,a}$,  G. Nardini$^{\,b}$, 
M. Quir\'os$^{\,b,c}$, C.E.M. Wagner$^{\,d,e}$}\\

\vspace{1.2cm}

${}^a\!\!$
{\em {Theoretical Physics Department, Fermilab, P.O. Box 500, Batavia, 
IL 60510}}

${}^b\!\!$
{\em { IFAE, Universitat Aut{\`o}noma de Barcelona,
08193 Bellaterra, Barcelona (Spain)}}

${}^c\!\!$
{\em {Instituci\`o Catalana de Recerca i Estudis Avan\c{c}ats (ICREA)}}

${}^d\!\!$
{\em {HEP Division, Argonne National Laboratory, Argonne, IL 60439}}

${}^e\!\!$
{\em {EFI, KICP and Physics Deparment, Univ. of Chicago, Chicago, IL 60637}}

\end{center}

\vspace{0.8cm}

\centerline{\bf Abstract}
\vspace{2 mm}
\begin{quote}\small

Electroweak baryogenesis in the minimal supersymmetric extension of
the Standard Model may be realized within the light stop scenario,
where the right-handed stop mass remains close to the top-quark mass
to allow for a sufficiently strong first order electroweak phase
transition.  All other supersymmetric scalars are much heavier to
comply with the present bounds on the Higgs mass and the electron and
neutron electric dipole moments.  Heavy third generation scalars
render it necessary to resum large logarithm contributions to perform
a trustable Higgs mass calculation. We have studied the one--loop RGE
improved effective theory below the heavy scalar mass scale and
obtained reliable values of the Higgs mass. Moreover, assuming a
common mass $\tm$ for all heavy scalar particles, and values of all
gaugino masses and the Higgsino mass parameter about the weak scale,
and imposing gauge coupling unification, a two-loop calculation yields
values of the mass $\tm$ in the interval between three TeV and six
hundred~TeV.  Furthermore for a stop mass around the top quark mass,
this translates into an upper bound on the Higgs mass of about
150~GeV.  The Higgs mass bound becomes even stronger, of about
129~GeV, for the range of stop and gaugino masses consistent with
electroweak baryogenesis. The collider phenomenology implications of
this scenario are discussed in some detail.

\end{quote}

\vfill

\newpage
\section{\sc Introduction}
\label{introduction}
The minimal supersymmetric extension of the Standard Model (MSSM) has
become the preferred candidate for the ultraviolet completion of the
Standard Model (SM) beyond the TeV scale.  The MSSM description may be
extended up to a high (GUT or Planck) scale, and the search for
supersymmetric particles is therefore one of the main experimental
goals at the forthcoming Large Hadron Collider (LHC) at CERN. Among
its main virtues, on top of solving the hierarchy problem of the
Standard Model, the MSSM leads to a natural unification of the gauge
couplings consistent with precision electroweak data and provides a
natural candidate for the Dark Matter of the Universe (namely the
lightest neutralino).

On the other hand electroweak baryogenesis~\cite{early} is a very
elegant mechanism for generating the baryon asymmetry of the Universe
that can be tested at present accelerator energies and, in particular,
at the future LHC. It turns out that electroweak baryogenesis can not
be realized within the Standard
Model~\cite{Shaposhnikov:1987tw,Gavela:1993ts}, while it is not a
generic feature of the MSSM for arbitrary values of its
parameters~\cite{Giudice:1992hh,Espinosa:1993yi}. However, a
particular region in the space of supersymmetric parameters was found
in the MSSM, where electroweak baryogenesis has a chance of being
successful~\cite{Carena:1996wj}, dubbed under the name of light stop
scenario (LSS). 

Since the generation of the BAU in the LSS is challenging other
alternatives (where the right-handed stop is not singled out) have
been explored in the literature. In particular in the context of split
supersymmetry, and if one allows $R_p$-violating couplings, it was
proven in Ref.~\cite{Huber:2005iz} that superheavy squarks can produce
enough baryon asymmetry when they decay out-of-equilibrium, while some
splitting between left and right-handed mass squarks is required by
the gluino cosmology. Moreover beyond the MSSM there are plenty of
other possibilities. The simplest one is introducing singlets in the
MSSM light spectrum (the so-called NMSSM~\cite{Pietroni:1992in} or
nMSSM~\cite{Menon:2004wv}), or even adding an extra $Z'$ gauge
boson~\cite{Kang:2004pp}, which easily triggers a strong first order
phase transition.

Since the generation of the BAU in the MSSM has inherent uncertainties
of order one, large variations in the final results appear due to the
different approaches which have been considered in the
literature~\cite{CQRVW}~\footnote{For instance, a possible contribution to
the baryon asymmetry coming from light sbottons and staus have been
recently explored in Ref.~\cite{Chung:2008ay}.}. According to these
results, it looks possible to achieve the proper baryon asymmetry
fulfilling all experimental bounds and in view of the forthcoming LHC
running, it is worth refining the predictions of the LSS.  In this
paper we will then consider the effective theory of the LSS while in a
companion paper~\cite{CNQW2} the phase transition will be analyzed in great
detail using the results provided by the present analysis.

The light stop scenario of the MSSM is characterized by a light
right-handed stop (with a mass near the top quark mass) while all
other squarks and sleptons should be heavy enough in order to cope
with present LEP bounds on the Higgs mass and to avoid large flavor,
CP violation and electric dipole moment effects~\cite{EDM,edm1}. On
the other hand, supersymmetric fermions (Higgsinos and gauginos) are
required to be at the electroweak (EW) scale (this fact can be
technically natural as a consequence of some partly conserved
$R$-symmetry) in order to trigger the required CP-violating currents
needed for baryogenesis~\cite{CQRVW}, as well as providing a Dark
Matter candidate~\cite{Balazs}. Moreover even if the LSS is consistent
with a light CP-odd Higgs boson, a large splitting between the
lightest CP-even Higgs and the CP-odd Higgs masses helps to avoid all
phenomenological constraints, because it emulates the Standard Model
Higgs sector at low energy (LE).

In practice we will consider all heavy scalars (sleptons, non-SM Higgs
bosons and squarks, except for the right-handed stop) at a common
scale $\tilde m$ and study the LE Effective Theory (ET) below that
scale. We will use the $\overline{MS}$ renormalization scheme and
resum the large logarithms which will appear in the calculation of
various observables by using Renormalization Group Equations (RGE)
techniques. In particular we will make use of the run-and-match
technique~\cite{Georgi:1994qn} by which every particle decouples at
its mass scale using the step-function approximation.  The high-energy
(HE) and LE theories, with different RGE in both regions, should match
at the decoupling scale providing (finite) thresholds for the various
couplings. In this way, considering a common decoupling scale is an
approximation which amounts to neglect possible thresholds
corresponding to the mass differences around $\tilde m$, and that
should not affect our results in a significant way.

For very large values of $\tilde m$ the model is a variant of Split
Supersymmetry~\cite{ArkaniHamed:2004fb}, where the right-handed stop
is also (light) in the LE theory. Thus in the spirit of Split
Supersymmetry every light particle is required by one particular
experimental input: apart from the light Higgs, required by
electroweak symmetry breaking, the light stop is required to trigger a
strong enough first order phase transition while light charginos and
neutralinos are required to generate enough baryon asymmetry and to
become dark matter candidates. On the other hand gauge coupling
unification, which works reasonably well in the MSSM, is an important
issue. As we will see a two-loop analysis points towards values of
$\tilde m$ between ten and one hundred TeV for the case where all
gauginos are at the electroweak scale, and around one order of
magnitude larger for hierarchical gaugino masses as required by
gaugino mass unification and by electroweak baryogenesis.

The outline of this paper is as follows. In Section~\ref{effective} we
present our LE effective theory below $\tilde m$ as well as the
matching conditions between the couplings of LE and HE theories, the
threshold conditions for the different couplings and the
$\beta$-functions in the LE Effective Theory. The technical details of
the calculation of threshold conditions are presented in
Appendix~\ref{thresholds} and those about the RGE in
Appendix~\ref{RGE}. In section~\ref{numerical} we present the
numerical results based on the calculation of the previous section. In
particular, the predictions of different parameters in the LE
effective theory and the corresponding value of the Higgs mass. In
Section~\ref{unification} we consider the issue of gauge coupling
unification.  We show that the unification scale is $M_{GUT}= 1\div2
\times 10^{16}$ GeV while imposing the experimental value for the
strong coupling leads to values of the heavy sfermion masses $\tm$ in
good agreement with the values of the parameters required to fulfill
the electric dipole moment constraints in the EWBG scenario within the
MSSM~\cite{EDM,edm1,Balazs}.  In Section~\ref{pheno} we present some
ideas for the experimental detection of $\tilde t_R$ in our model, as
well as the possibility of having a Dark Matter candidate. Finally in
Section~\ref{conclusion} we present our conclusions.

\section{\sc The effective theory}
\label{effective}

The theory at an energy scale $\tau$ between the EW scale and $\tm$,
at which supersymmetry is broken, contains all the SM particles and
the Bino, Winos and Higgsinos, as well as the light stop. All other
squarks and sleptons are heavy, with masses about $\tm$, and decouple
from the low energy theory.  The gluino, with a mass $M_3$ much below
$\tm$, may be much heavier than the other gauginos and, in this case,
when $\tau < M_3$ it will decouple too.  Therefore the corresponding
low energy effective Lagrangian is given by
 
\bea
{\cal L}_{\textrm eff} &=& m^2 H^\dagger H-\frac{\lambda}{2}\left(
H^\dagger H\right)^2 - h_t \left[{\bar q}_L \epsilon H^* t_R \right] +
Y_t \left[\overline{{\tilde H}}_u\epsilon q_L {\tilde t_R}^*
\right] 
\nonumber \\ 
&&-\frac{M_3}{2}\Theta_{\tilde g}\, {\tilde g}^a {\tilde g}^a -\frac{M_2}{2} {\tilde W}^A
{\tilde W}^A -\frac{M_1}{2} {\tilde B} {\tilde B} -\mu {\tilde
H}_u^T\epsilon {\tilde H}_d  - M_U^2 \,\left|{\tilde t_R}\right|^2
\nonumber \\  && - \sqrt{2} \Theta_{\tilde g} G~ {\tilde t_R} {\tilde g}^a
\overline{T}^a \overline t_R  + \sqrt{2} J~ {\tilde t_R} {\tilde B}
\overline t_R
-\frac{1}{6} K \left|{\tilde t_R}\right|^2 \left|{\tilde t_R}\right|^2
 - Q ~ \left|{\tilde t_R}\right|^2
\left|{H}\right|^2 ~+{h.c.}
\nonumber \\
&& +\frac{H^\dagger}{\sqrt{2}}\left( g_u \sigma^a {\tilde W}^a + g'_u
{\tilde B} \right) {\tilde H}_u +\frac{H^T\epsilon}{\sqrt{2}}\left( -
g_d \sigma^a {\tilde W}^a + g'_d {\tilde B} \right) {\tilde H}_d
+{\textrm {h.c.}} ~ ,
\label{lagreff}
\eea 
where the gluino decoupling is taken into account by the symbol
$\Theta_{\tilde g}$ which is equal to 1 (0) for $\tau\geq M_3 (\tau<
M_3)$. For simplicity in (\ref{lagreff}) we do not write the kinetic
terms explicitly and we approximate the Lagrangian by taking into
account only interactions of the SM fields, charginos, neutralinos and
the right-handed stop coming from renormalizable high energy terms
proportional to the gauge couplings $g', g, g_3$ or the supersymmetric
top Yukawa coupling $\lambda_t$ without considering flavour mixing.
Furthermore in (\ref{lagreff}) the field $H$ is defined as the light
projection of the MSSM Higgs bosons, given by
$H_u \to \sbe H\,,H_{d,i} \to \cbe \epsilon_{ij} H^*_j$, with
$\tbe\equiv \langle H_u^0 \rangle/\langle H_d^0 \rangle$.

At the energy scale $\tm$ the effective Lagrangian (\ref{lagreff})
has to describe the physics of the HE theory, which implies that the
following one--loop {\it matching conditions} have to be satisfied
\bea
\label{matchQ}
Q(\tilde m )-\Delta Q&=&\left(\lambda_t^2 (\tilde m )\sin^2\beta +
      \frac{1}{3}~g'^2(\tm) \cos 2\beta\right) \left(1-{1\over2} \Delta
      Z_Q\right)~,\\
\label{matchLamb}
\lambda(\tilde m ) -\Delta\lambda&=& \frac{g^2(\tilde m )+g^{\prime
      2}(\tilde m)} {4} \cos^22\beta \left(1-{1\over2}\Delta
      Z_\lambda\right)~, \label{lambda}\\ K (\tilde m )-\Delta K &=&
      \left(g_3^2
      (\tm)+\frac{4}{3}~g'^2(\tm)\right)\left(1-{1\over2}\Delta
      Z_K\right) ,\label{matchK}\\
G (\tilde m )-\Delta G&=&g_3 (\tilde m ) \left(1-{1\over2}\Delta
Z_G\right)~,\\
h_t(\tilde m )-\Delta {h_t}&=&\lambda_t (\tilde m )\sin\beta
\left(1-{1\over2}\Delta Z_{h_t}\right)~,
\label{matchht}
\eea
\bea
Y_t (\tilde m )-\Delta {Y_t}&=& \lambda_t (\tilde m
 )\left(1-{1\over2}\Delta Z_{Y_t}\right)~, \label{matchY}\\
g_u(\tm)=g(\tm)\sin\beta~, && g_d(\tm)=g(\tm)\cos\beta ~,
\label{weak1} \\
g'_u(\tm)=g'(\tm)\sin\beta~, && g'_d(\tm)=g'(\tm)\cos\beta
~,\label{weak2}\\
J (\tilde m )=\frac{2}{3}g'(\tilde m ) ~,&& \label{match}
\eea
where the quantities $\Delta Q$, $\Delta \lambda$, $\Delta
K$, $\Delta G$, $\Delta h_t$, $\Delta Y_t$ and $\Delta Z_i$ are the
{\it threshold} functions. In particular $\Delta Z_i$ are the wave
function thresholds coming from the matching of low and high energy
propagators and the canonical normalization of ET kinetic terms while
the others come directly from the matching of the low and high energy
proper vertices (details of the calculation are given in
Appendix~\ref{thresholds}).

In this work we will consider for the threshold and $\beta$-functions
the leading contributions and thus we will use the approximation of
neglecting the one--loop corrections proportional to $g'$, $g$ and the
Yukawa couplings other than that of the top-quark (as well as the low
energy couplings correlated to those). Following this criterion we
consider no threshold in the matchings (\ref{weak1})-(\ref{match})
since they do not appear at tree-level and would correspond to the
one--loop corrections that we are neglecting.

The same analysis has to be redone when the renormalization scale
$\tau$ becomes lower than $M_3$ and the gluino decouples. In this case
the interaction term of (\ref{lagreff}) involving the coupling $G$
disappears and the following matching conditions relate the values of
the couplings before and after the gluino decoupling:
\bea
\label{matchgl}
Q(M_3^-)&=&Q(M_3^+) (1-\Delta' Z_\str) ~, \nonumber \\
K(M_3^-)&=&K(M_3^+) (1-2 \Delta' Z_\str) + \Delta' K~,\nonumber\\
h_t(M_3^-)&=& h_t(M_3^+) (1-\Delta' Z_{t_R} /2) ~,\\
Y_t(M_3^-)&=& Y_t(M_3^+) (1-\Delta' Z_\str /2) ~, \nonumber \\
M_U^2(M_3^-)&=&M_U^2(M_3^+) (1-2 \Delta' Z_\str) + \Delta' M_U^2~,\nonumber
\eea
where $\Delta' Z_\str$ ($\Delta' Z_{t_R}$) is the wave function
threshold of the right stop (top) and $\Delta' K$ and $\Delta' M_U^2$
are the proper vertex threshold. The matching conditions of the
couplings absent from (\ref{matchgl}) are trivial since they have no
threshold discontinuity when the renormalization scale crosses $M_3$.
Readers interested in the explicit form of the thresholds of
(\ref{matchQ})-(\ref{matchgl}) can find them in
Appendix~\ref{thresholds}, Eqs.~(\ref{deltaQ})-(\ref{wfgl}) and
(\ref{thresholdM3}).

For energy scales between the top mass and $\tilde{m}$, at which all
scalars apart from the right-handed stop and the Standard Higgs
doublet are decoupled, one can compute the one-loop $\beta$-functions
of the gauge constants~\footnote{The two-loop beta functions will be
given in Section~\ref{unification} where the issue of gauge coupling
unification is considered.} in a straightforward way
\bea
(4\pi)^2 \beta_{g_i} = g_i^3 b_i      
	~~~~\mathrm{with} ~~~b=\left(
		\frac{143}{30} , -\frac{7}{6} ,
                -\frac{41}{6}+2 \Theta_{\tilde g}  \right) 
		\nonumber~,
\eea
where we have used the GUT convention $g_1^2=(5/3)g'^{2}$.  

For the RGE of the other couplings we will only report their
expressions and leave the calculation details to Appendix~\ref{RGE}.
For the dimensionless couplings we obtain
\bea
\label{stopbeta}
        &&(4\pi)^2 \beta_{g_u}= g_u 
                \left(3 h_t^2 + \frac{3}{2} Y_t^2\right) ~,
	~~ (4\pi)^2 \beta_{g_d}= 3 ~g_d~ h_t^2 ~,\nonumber \\
	&&(4\pi)^2 \beta_{g'_u}= g'_u 
                \left(3 h_t^2 + \frac{3}{2} Y_t^2\right) ~, 
	~~(4\pi)^2 \beta_{g'_d}= 3 ~g'_d~ h_t^2 ~, \nonumber \\
        &&(4\pi)^2 \beta_{J}= ~J~\left(h_t^2+2
        Y_t^2+\frac{12}{3}~G^2 \Theta_{\tilde g} -4g_3^2\right) ~,\nonumber \\ 
	&&(4\pi)^2 \beta_{Y_t}= \frac{1}{2}Y_t \left(
		h_t^2 + 8 Y_t^2 + \frac{16}{3} G^2 \Theta_{\tilde g}-8g_3^2
		\right) ~, \nonumber \\
	&&(4\pi)^2 \beta_G = \frac{1}{2}G \left(
		9G^2 + 2h_t^2 - 26 g_3^2 + 4 Y_t^2
		\right) ~, \\
	&&(4\pi)^2 \beta_{h_t} = h_t \left(
	\frac{9}{2}h_t^2 + \frac{1}{2}Y_t^2 + \frac{4}{3}G^2 \Theta_{\tilde g}
		-8 g_3^2\right) ~, \nonumber \\
	&&(4\pi)^2 \beta_\lambda = 12 \lambda^2 +6 Q^2 -12 h_t^4 
		+12 h_t^2 \lambda ~,\nonumber \\
	&&(4\pi)^2 \beta_Q = 
		- \frac{32}{3} G^2 h_t^2  \Theta_{\tilde g}- 4 Y_t^2 h_t^2 
		+Q \left(K + 3 \lambda + 4 Q+ 6 h_t^2 + 4 Y_t^2 +
		\frac{16}{3} G^2  \Theta_{\tilde g}- 8 g_3^2
			\right) ~, \nonumber \\
	&&(4\pi)^2 \beta_K = 12 Q^2 +13 g_3^4
		- \frac{88}{3}G^4  \Theta_{\tilde g}-24 Y_t^4 
		+ K \left(\frac{14}{3}K +
			8 Y_t^2 + \frac{32}{3} G^2  \Theta_{\tilde g}
                - 16 g_3^2 
		\right)  \nonumber ~,
\eea
and for the dimensionful ones
\bea
\label{massbeta}
        &&(4\pi)^2 \beta_{\mu}= \frac{3}{2}~\mu~ Y_t^2~, \nonumber \\
        &&(4\pi)^2 \beta_{M_1}= {\cal O}(g_1^2)~,
	~~ (4\pi)^2 \beta_{M_2}= {\cal O}(g_2^2) ~, \nonumber \\
	&&(4\pi)^2 \beta_{M_3}= M_3
                \left(-18 g_3^2 + G^2\right)  ~, \\
	&&(4\pi)^2 \beta_{m}= -6 Q~M_U^2 +
            6 m^2  h_t^2 ~, \nonumber \\
	&&(4\pi)^2 \beta_{M_U^2}= M_U^2 \left(
		 \frac{8}{3} K + 4 Y_t^2 + \frac{16}{3}G^2 -8 g_3^2
		\right)-\frac{32}{3} M_3^2 G^2 \Theta_{\tilde g} -4 m^2 Q -4 Y_t^2 \mu^2
		 ~,
 \nonumber
\eea
where $\beta_G$ and $\beta_{M_3}$ make sense only for $\tau\geq M_3$.

\section{\sc Numerical results on the Higgs mass}
\label{numerical}
In this section we will apply the previous results to obtain in an
appropriate way the values of the LE couplings and the Higgs mass at
the EW scale for any large value of the cutoff scale $\tilde m$ and
for different values of the HE supersymmetric parameters.

\subsection{\sc Running of couplings}

We need to know all the couplings of (\ref{lagreff}) at the EW scale
that we identify here with the top-quark mass $m_t=172.5\pm 2.7$
GeV~\cite{Yao:2006px} (corresponding to $h_t(m_t)\simeq 0.95$).  All
the mass parameters $M_U^2,\mu,M_3$ are free inputs of the theory and
thus we choose them directly at low energy by fixing $M_U^2(m_t),
\mu(m_t),M_3(M_3)$. Moreover at the low scale also the SM couplings
$g(\tm), g'(\tm), g_3(\tm),h_t(\tm)$ and $m^2(m_t)$ are fixed
experimentally~\footnote{The parameter $m^2(m_t)$ is fixed by the
condition that the minimum of the SM-like Higgs one--loop potential be
$v=246.22$ GeV at the scale $m_t$.}. On the contrary the non-SM
couplings are defined by (\ref{matchQ})-(\ref{match}) at high energy
as functions of the previous couplings, run up to the scale $\tm$, and
the free quantities $\tm, \tbe$ and $A_t(\tm)$. Therefore in order to
get the non-SM couplings at the EW scale we have to solve a system of
linear differential equations [the RGE
(\ref{stopbeta})-(\ref{massbeta})] with boundary condition in
$\tau=m_t,M_3,\tm$. Equations must be solved numerically and
iteratively because the conditions at the boundary $\tm$
(\ref{matchQ})-(\ref{match}) depend in turn on the evolution of the
parameters.  The implicit resummation of the leading logarithms
renders our estimation of the ET couplings reliable, even for large
values of $\tm$. Using this procedure the values of the
\vspace{4mm}
\begin{figure}[htb]
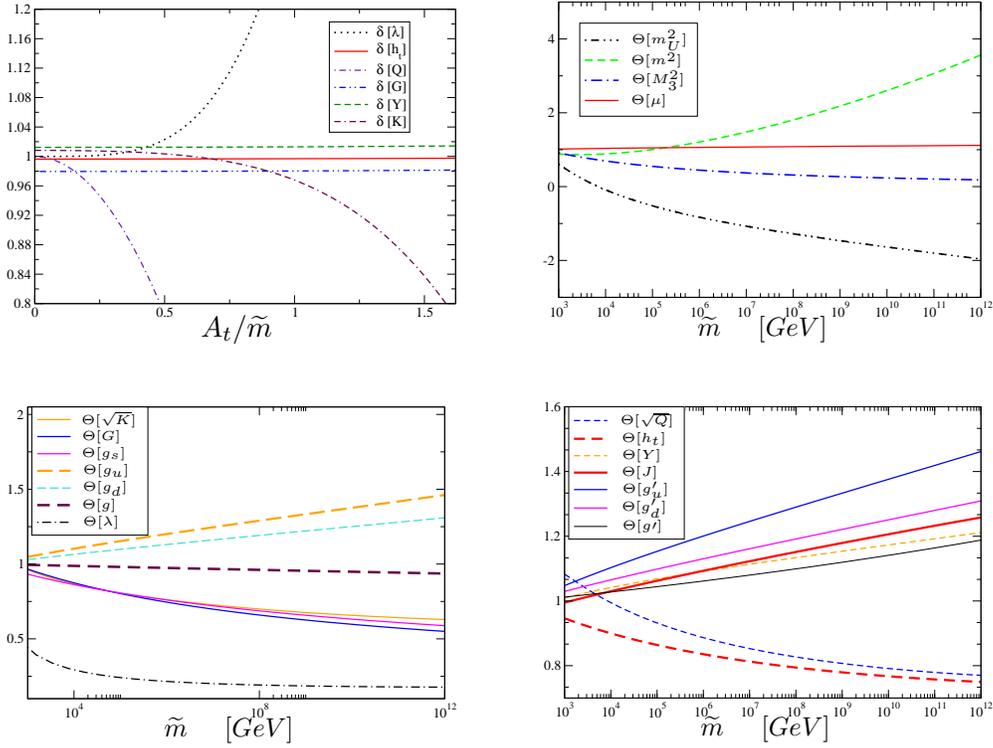

\psfrag{XX}[][bl]{$A_t/\tm$}
\psfrag{X1}[][bl]{\Large${}_{\tm ~~~{[GeV]}}$}

\psfrag{KKK}[][c]{\tiny ${}_{~~~~~\Theta[\sqrt{K}]}$}
\psfrag{GGGGGGG}[][c]{\tiny ${}_{\hspace{-2.6mm}\Theta [G]}$}
\psfrag{GSSSSS}[][c]{\tiny ${}_{~\Theta [g_s]}$}
\psfrag{GUUUUU}[][c]{\tiny ${}_{\Theta [g_u]}$}
\psfrag{GDD}[][c]{\tiny ${}_{~~~~\Theta [g_d]}$}
\psfrag{gggg}[][c]{\tiny ${}_{~~\,\Theta [g]}$}
\psfrag{lamb}[][c]{\tiny  ${}_{~~~\Theta [\lambda]}$}

\psfrag{QQQQQQQ}[][c]{\tiny ${}_{\Theta[\sqrt{Q}]}$}
\psfrag{hhhhh}[][c]{\tiny ${}_{~~~~\Theta[h_t]}$}
\psfrag{YYYY}[][c]{\tiny ${}_{~~\,\Theta[Y]}$}
\psfrag{JJJJJ}[][c]{\tiny ${}_{~~~\;\Theta[J]}$}
\psfrag{guuuuu}[][c]{\tiny ${}_{~~~\,\Theta [g_u']}$}
\psfrag{gdddd}[][c]{\tiny  ${~~~~}_{\Theta [g'_d]}$}
\psfrag{gygyg}[][c]{\tiny  ${}_{~~~\,\Theta[g\prime]}$}
%
\psfrag{mUUUUUU}[][c]{\tiny ${}_{\Theta [m^2_U]}$}
\psfrag{mmmmmm}[][c]{\tiny ${}_{\Theta[m^2]}$}
\psfrag{M333}[][c]{~~~\,\tiny ${}_{\Theta[M^2_3]}$}
\psfrag{muuuu}[][c]{~\tiny ${}_{\Theta[\mu]}$}
\begin{center}
\epsfig{file=thresholdsTbeta2.eps,width=0.4\textwidth}\hspace{1cm}
\epsfig{file=running3.eps,width=0.4\textwidth} \\[.9cm]
\epsfig{file=running1.eps,width=0.4\textwidth}
\hspace{1cm}\epsfig{file=running2.eps,width=0.4\textwidth}
\end{center}
\caption{Upper--left panel: We plot for every coupling the ratio
  $\delta$ of its value at the decoupling scale $\tm$ over its value
  without any threshold contribution, as a function of $A_t/\tm$ for
  $\tm = 100$~TeV.  Upper-right and lower panels: the ratio $\Theta$
  between the couplings at $\tau=m_t$ and their starting value at
  $\tau=\tm$ (for the coupling $M_3$ and $G$ the lower $\tau$ value is
  $\tau=M_3$) is plotted as function of $\tm$, for $A_t=0.6~\tm$. In
  all plots $\tbe=2$, $M_U=200$~GeV, $M_3=500$~GeV and $\mu=100$~GeV
  have been fixed. }
\label{PlotThresh}
\end{figure}
 LE couplings at the EW scale
will basically depend on two different factors: the matching
conditions and the running evolution. Focusing on the former in the
upper--left panel of Fig.~\ref{PlotThresh} we analyze the thresholds
relevance by plotting for every coupling the ratio (defined as
$\delta$ in the plot) of its value over the one without the threshold
contribution, both evaluated at the scale $\tm$, as functions of
$A_t/\tm$ for $\tbe=2, M_U=200\, $GeV$, M_3=500~ $GeV$, \mu=100~ $GeV,
and $\tm=100~ $TeV. It is remarkable that the threshold contributions
to the couplings $\lambda, Q$ and $K$ can easily reach a value $\sim
10\%$ and beyond, unlike the $h_t, Y, G$-thresholds which are almost
$A_t$-independent and remain below $\sim 2\%$~\footnote{It has been
checked that this estimate holds also for other values of $\tm$.}.
For this reason it is sensible to neglect the threshold effects of
$h_t, Y, G $, since their contributions are within the uncertainty of
our approximations.

The relevance of the running is also exhibited in
Fig.~\ref{PlotThresh} where the ratios $\rho(\tm)/\rho(X)\equiv
\Theta[\rho]$ for all couplings $\rho$ are plotted as functions of
$\tm$ (where $X=M_3$ for $\rho=M_3,\, G$ and $X=m_t$ for the rest of
couplings) for $A_t=0.6\,\tm$ and keeping the rest of parameters fixed as
in the upper--left plot.  In particular in the upper--right panel we
plot masses and in the lower panels all dimensionless couplings. For
example we can compare from the lower--right figure how the couplings
$h_t$ and $Y$ evolve differently, even if we had neglected their
different threshold effects.

\subsection{\sc The Higgs mass}

Once we have computed the values of the couplings in (\ref{lagreff})
it is straightforward to obtain the Higgs effective potential in which
the leading logarithms are resummed. Since this potential is strongly
dependent on the renormalization scale we need to consider the one--loop
part of the effective potential calculated in the LE theory. We are
adding to the SM fields only the contribution from $\tilde t_R$ since
the contribution from charginos and neutralinos (which is numerically
small) would spoil the scale invariance of the effective potential in
our approximation where we are neglecting electroweak gauge couplings
in the LE $\beta$-functions. The one--loop contributions to the
effective potential then read as
\bea
 V_{1-loop}(\phi_c)= 
\label{potForm}
 \frac{6}{64 \pi^2} \sum_i
           n_i~m_i^4(\phi_c) \left(
                   \ln \frac{m_i^2(\phi_c)}{\mu^2}
                             -C_i\right)  
	\quad \textrm {with} ~~~{i=W,Z,h,\chi,\str, t} 
\eea
\begin{figure}[htb]
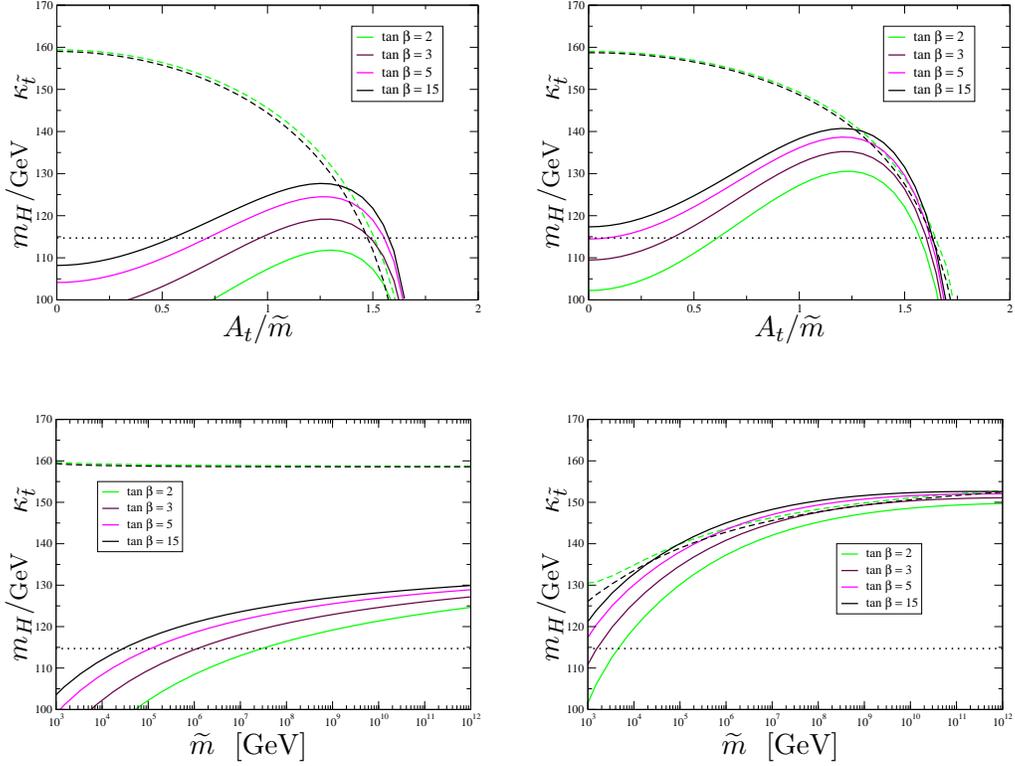

\psfrag{Y1}[][bl]{$m_H/\textrm {\footnotesize GeV}  ~~~~ \kappa_{\tilde t}$}
\psfrag{Y2}[][bl]{$m_H/\textrm {\footnotesize GeV}  ~~ ~~
  \kappa_{\tilde t} $}
\psfrag{X1}[][l]{$A_t/\tm$}
\psfrag{X2}[][l]{$\tm$~~[GeV] }
\vspace{.5cm}
\begin{center}
\epsfig{file=3TevM200.eps,width=0.42\textwidth}\qquad
\epsfig{file=100TevM200.eps,width=0.42\textwidth}
\\ \vspace{10mm}
\epsfig{file=at0M200.eps,width=0.42\textwidth}\qquad
\epsfig{file=at13M200.eps,width=0.42\textwidth}
\end{center}
\caption{$m_H/\textrm{GeV}$ (solid lines) and $\kappa_{\tilde t}$
(dashed curves) predicted for the case $M_U=200$ GeV, $\mu=100$ GeV
and $M_3=500$ GeV for several values of $\tbe$.  In the upper--left
(upper--right) panel $\tm=3$ (100) TeV has been fixed and in the
lower--left (lower--right) panel $A_t=0 (1.3) \tm$.  The experimental
lower bound on the Higgs mass is marked by a dotted straight line.
\label{plotStH}}
\end{figure}
\noindent where $C_W=C_Z=5/6$, $C_h=C_\chi=C_\str=C_t=3/2$ and
$n_W=n_\str=6,\,n_Z=3,\,$  $n_h=1,\,n_\chi=3,\,n_t=-12$
and the masses are
\bea 
m_W^2=\frac{g^2}{4}\phi_c^2 ~, &&
m_Z^2=\frac{g^2+g'^2}{4}\phi_c^2 ~, 
\nonumber\\
m_h^2=\frac{\lambda}{2}(3\phi_c^2-v^2) ~, &&
m_\chi^2=\frac{\lambda}{2}(\phi_c^2-v^2) ~, 
\nonumber\\
 m_\str^2=M_U^2 + \frac{Q}{2}\phi_c^2 ~, &&
 m_t^2=\frac{h_t^2}{2} \phi_c^2 ~,  \label{stopmass}
\eea
with the renormalization scale conventionally chosen to be $m_t$.
Notice that by this renormalization scale choice and thanks to the use
of the LE theory the logarithms of (\ref{potForm}) are always small.
Moreover the addition of the one--loop contribution (\ref{potForm})
eliminates the scale dependence of the potential proportional to
strong--like or Yukawa--like couplings up to the one--loop order.

The second derivative of the potential at the EW minimum provides the
Higgs mass within the one-loop renormalization group improved
effective theory.  The numerical result is shown in Fig.~\ref{plotStH}
where we plot the Higgs mass $m_H$ (solid line). We also introduce the
parameter $\kappa_{\tilde t}\equiv 10 \sqrt{m_{\tilde t_R}/
\textrm{GeV}}$ (dashed line) which parameterizes the lightest stop
mass. The parameter $\kappa_{\tilde t}$ has the advantage of being
related in a simple way to the stop mass, and since it acquires values
similar to the Higgs mass (in GeV units), it may be represented
together with it on a linear scale. Observe that $\kappa_{\tilde t} =
100$ is equivalent to $m_{\tilde t_R} = 100$~GeV and $m_{\tilde t_R} <
m_t$ corresponds to $\kappa_{\tilde t} \lesssim 130$. Values of
$\kappa_{\tilde t} \lesssim 100$ are therefore excluded by LEP
searches.  We plot both variables as functions of $A_t/\tm$ [upper
panels: on the left (right) panel $\tm=3$ (100) TeV] and $\tm$ [lower
panels: on the left (right) panel $A_t=0\, (1.3)\, \tm$] for several
values of $\tbe$.

For $m_H$ and $\kappa_{\tilde t}$ the different values of $\tbe$ are
encoded by different colours (level of line darkness) presented in the
legend. Since a change of $\tbe$ does not appreciably modify
$\kappa_{\tilde t}$ we mark only the extremal curves corresponding to
$\tbe=15$ and $\tbe=2$. In all the plots we have fixed $M_U=200$ GeV,
$\mu=100$ GeV and $M_3=500$ GeV.

Some comments on the different masses can be easily drawn from
Fig.~\ref{plotStH}. We notice that because of the experimental bound
on the Higgs mass, $m_H > 114.7$ GeV~\cite{Yao:2006px} (dotted
straight line in Fig.~\ref{plotStH}) the model with $\tm \sim 1$ TeV
requires $\tbe> 3$ and in general the smaller $\tbe$ is the closer to
1.3$\,\tm$ the triliniear coupling $A_t$ has to be since the Higgs
mass has a maximum there. This requirement is relaxed if $\tm$ is
increased, but scales as large as $\tm\simeq 10^7$ TeV are necessary
to overcome the Higgs mass bound for any $\tbe\gtrsim 2$ independently
of $A_t$. On the contrary if we allow $A_t\simeq 1.3\, \tm$ the model
is experimentally safe for $\tbe\geq 2$ already at $\tm=5$ TeV.  On
the other hand if we require the right-handed stop to be lighter than
the top quark ($\kappa\lesssim 130$) with $M_U^2\sim (200~
\textrm{GeV})^2$ a large $A_t$ is needed.  Another way to maintain the
stop lighter than the top quark is by decreasing $M_U^2$, which lowers
the stop mass.  For $M_U^2\lesssim (100~ \textrm{GeV})^2$ the maxima
of the Higgs mass curves are excluded by the LEP bound on the stop
mass. Consequently the bounds on $\tbe$, $A_t$ and $\tm$ become even
stronger in this case.

\begin{figure}[h]
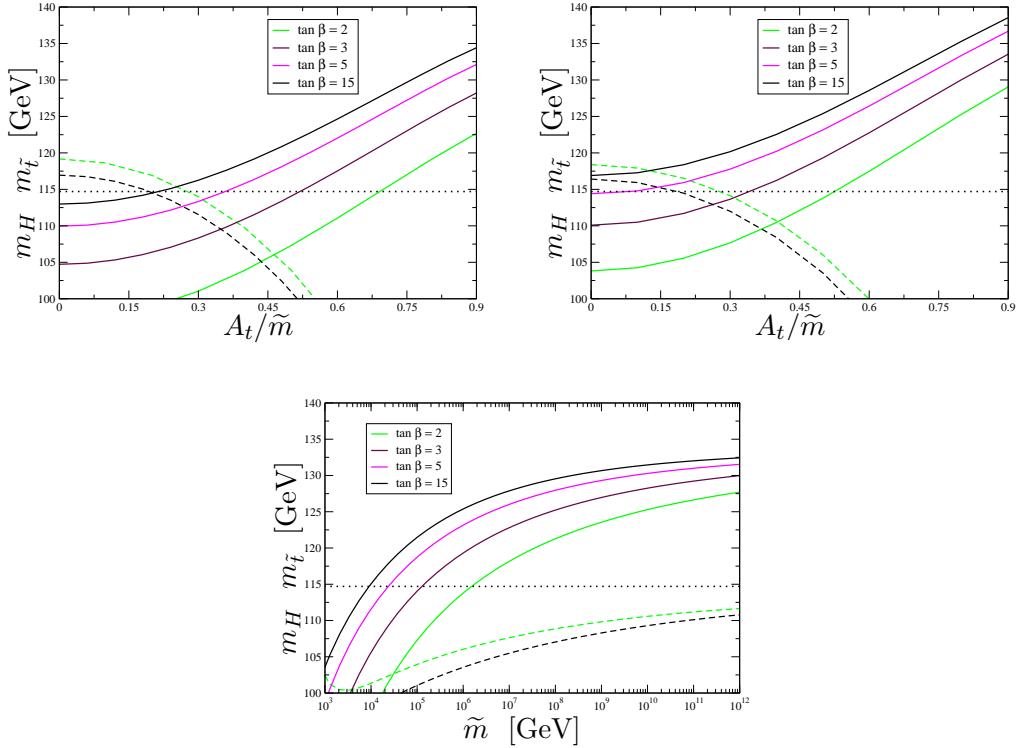

\psfrag{Y1}[][bl]{$m_H  ~~ m_{\tilde t} ~~[\textrm{GeV}]$}
\psfrag{Y2}[][bl]{$m_H  ~~ m_{\tilde t} ~~[\textrm{GeV}]$}
\psfrag{X1}[][l]{$A_t/\tm$}
\psfrag{X2}[][l]{$\tm$~~[GeV]}

\vspace{.5cm}
\begin{center}
\epsfig{file=100TevStop_100.eps,width=0.42\textwidth}\qquad
\epsfig{file=1000TevStop_100.eps,width=0.42\textwidth}\\[8mm]
\epsfig{file=at6M_100.eps,width=0.42\textwidth}
\end{center}
\caption{Plots similar to those in upper panels of Fig.~\ref{plotStH},
but with a different value of $M_U$ [$M_U^2=-(100\, {\rm GeV})^2$] and
of $\tm$ ($\tm=100$ TeV in the left panel and $\tm=1000$ TeV in the
right one). The lower plot is done with $A_t/\tm=0.5$ .
\label{plot100}}
\end{figure}

The latter result applies, in particular, for the conditions which are
favorable to electroweak baryogenesis (EWBG) where $M_U^2<0$ is
needed. As an example, in Fig.~\ref{plot100} we choose the same
parameters as in Fig.~\ref{plotStH} but with a right-handed stop mass
parameter $M_U^2 = -(100$~GeV$)^2$ and $\tm=100$ (1000) TeV in the
upper--left (right) plot. We can see from the right--panel of
Fig.~\ref{plot100} that there exists the upper bound $A_t \lesssim 0.6
\; \tm$ coming from the experimental bounds on the stop mass. Notice
that, independently of the experimental bounds, larger values of
$A_t/\tm$ would lead to an instability of the electroweak minimum.
Moreover values of $A_t/\tm \lesssim 0.5$ are also required in order to
obtain a strong enough electroweak phase
transition~\cite{Carena:1996wj}. On the other hand the rough estimate
$\tbe\lesssim 10$~\cite{EDM,Balazs}, coming from the requirement of
generation of the observed baryon asymmetry of the Universe, pushes
the parameter $\tm$ towards values $\tm\gg 1$~TeV which justifies a
posteriori the study of the effective theory with resummed logarithms.
A detailed analysis of electroweak baryogenesis in the present model
will be thoroughly analyzed in Ref.~\cite{CNQW2}.  Let us stress that
values of $\tm \gg 1$ TeV are consistent with those necessary in order
to suppress the one-loop contributions to the electric dipole moment
of the electron and the neutron in the light stop
scenario~\cite{edm1}.

Finally let us observe that all previous comments, which apply for a
gluino mass of 500 GeV, can also be extended to other values of the
gluino masses. In particular we have checked that for $M_3\simeq 1$
TeV the Higgs mass only decreases by a few percent with respect to the
case of gluino masses at the EW scale.

\section{\sc Gauge coupling unification}
\label{unification}

In this section we will consider the issue of gauge coupling
unification in the theory where below the scale $\tm$ there is the ET
which has been considered in Section~\ref{effective} and beyond $\tm$
the MSSM. In the extreme case where $\tm$ is at the EW scale, the
condition of gauge coupling unification yields low energy values for
the strong gauge coupling $\alpha_3$ consistent with those obtained in
low energy MSSM scenarios.  The MSSM prediction, however, depends
strongly on the possible threshold corrections to the gauge couplings
at the GUT scale, as well as on the additional threshold corrections
induced by the weak scale supersymmetric particles.  Ignoring
high-energy threshold corrections, one obtains a range of values
$\alpha_3(M_Z) =$ 0.120--0.135, with the exact value depending on the
precise MSSM spectrum. This range of values is compatible with
experimental data, but with some tension towards a predicted high
value.  When $\tm$ is increased the predicted value of $\alpha_3(M_Z)$
coming from the requirement of gauge coupling unification moves
towards lower values.  Therefore for a given low energy spectrum one
can find agreement with the experimental values for a certain range of
values of $\tm$.  In this sense it is possible to make a grand
unification ``prediction'' for the parameter $\tm$. High energy
threshold corrections would lead to an uncertainty on this range of
$\tm$ values. In this section, we will quantify these issues after
considering the two-loop RG evolution of the gauge couplings.

The two-loop renormalization-group equation for the gauge couplings
are~\cite{Machacek:1983tz}
\bea
&&(4\pi)^2\frac{d}{dt} g_i=g_i^3b_i \nonumber\\
&&+\frac{g_i^3}{(4\pi)^2} \left[
\sum_{j=1}^3B_{ij}g_j^2-d_i^u h_t^2 -d_i^W\left(\gtiluq +\gtildq \right)
-d_i^B\left( \gtilupq +\gtildpq \right) -d_i^G G^2 -d_i^J J^2 \right] 
\label{gauger}
\eea
where $t=\ln \tau$, $\tau$ is the renormalization scale and we use the
convention $g_1^2=(5/3)g^{\prime 2}$. Eq.~(\ref{gauger}) is
scheme-independent up to the two-loop order.
 
In the effective theory below $\tilde{m}$, the $\beta$-function
coefficients are
\bea
&&b=\left(\frac{143}{30},-\frac{7}{6},-\frac{41}{6}+2\Theta_{\tilde g}\right) 
~,~~~~B=
\begin{pmatrix}
\frac{376}{75}&\frac{18}{5}&\frac{196}{15}\\ 
\frac{6}{5} & \frac{106}{3}&12 \\
\frac{49}{30}&\frac{9}{2}&-\frac{67}{3}+48\Theta_{\tilde g} 
\end{pmatrix} \,, \\
&&d^u=\left(\frac{17}{10},\frac{3}{2},2\right) ~,~~~
d^G=\left(\frac{32}{10},0,\frac{13}{2}\right)\Theta_{\tilde g} ~,~~~
d^J=\left(\frac{18}{5},0,1\right) \,, \\
&&d^W=\left(\frac{9}{20},\frac{11}{4},0\right) ~,~~~
d^B=\left(\frac{3}{20},\frac{1}{4},0\right) \,,
\eea
while above $\tm$ one has the MSSM result [replacing in (\ref{gauger})
the SM Yukawa $h_t$ by the MSSM one $\lambda_t$ related to the former
by (\ref{matchht})]~\cite{Martin:1993zk}
\bea
&&b=\left(\frac{33}{5},1,-3\right) ~,~~~~B=
\begin{pmatrix}
\frac{199}{25}&\frac{27}{5}&\frac{88}{5}\\ 
\frac{9}{5} & 25&24 \\
\frac{11}{5}&9&14 
\end{pmatrix} \,, \\
&&d^u=\left(\frac{26}{5},6,4\right) ~,~~~
d^G=d^J=d^W=d^B=0 \,.
\eea
Finally the one-loop RGE of the Yukawa--like and gauge--like couplings
are given in Eq.~(\ref{stopbeta}), while for the supersymmetric Yukawa
coupling~\cite{Martin:1993zk}
\bea
(4\pi)^2 \frac{d \lambda_t}{dt}= \lambda_t 
  \left[-\frac{13}{15}g_1^2 -3 g_2^2 -\frac{16}{3} g_3^2 +6 \lambda_t^2\right]
  ~.
\eea

\begin{figure}[htb]
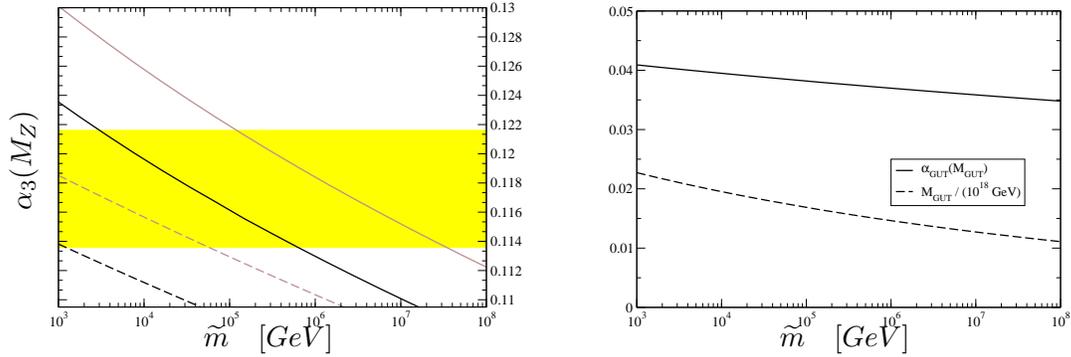

\psfrag{y}[][bl]{$\alpha_3(M_Z)$}
\psfrag{x}[][bl]{$\tm ~~~[GeV]$}
\psfrag{X}[][bl]{$\tm ~~~[GeV]$}

\vspace{.5cm}
\begin{center}
\epsfig{file=figUnif.eps,width=0.45\textwidth}
\hspace{1cm}\epsfig{file=figAlphaGut.eps,width=0.41\textwidth}
\end{center}
\caption{Predicted values of $\alpha_3(M_Z)$ [left panel: black lines
    are for all gauginos at the EW scale and grey lines for all
    gauginos at the EW scale except for the gluino with a mass
    $M_3=500$ GeV] by means of the two-loop (solid line) and one-loop
    (dashed-line) RG evolution, and the resulting two-loop predictions
    for $M_{GUT}$ and $\alpha_{GUT}$ [right panel] as functions of
    $\tm$ from the unification condition. The (yellow) band shows the
    experimental value of $\alpha_3(M_Z)$ within $2\,\sigma$.}
\label{unif}
\end{figure}

We will consider the following experimental inputs~\cite{Yao:2006px}
\bea
&&\sin^2\theta_{\overline{MS}}(M_Z)=0.2312\pm 0.0002 ~,\\
&& \alpha^{-1}_{EM}(M_Z)=127.906\pm 0.019 ~,\\
&&\alpha_3(M_Z)=0.1176\pm 0.0020 ~,
\eea
and by imposing unification of $\alpha_1(M_{GUT})=\alpha_2(M_{GUT})$
we obtain a prediction for $\alpha_3(M_Z)$ as it is shown in the left
panel of Fig.~\ref{unif}. The solid black line in the left panel of
Fig.~\ref{unif}
represents the two-loop result for values of all
gaugino masses about the weak scale, while the dashed black line
represents the one-loop result. The grey lines are corresponding plots
for a gluino mass $M_3=500$ GeV, which roughly follows the gaugino
mass unification relation $M_3/M_2 \simeq 3$. In the figure the
experimental value of $\alpha_3(M_Z)$ within $2\,\sigma$ is marked by
a (yellow) band. For our two choices of $M_3$ the gluino decoupling
almost does not modify the curves of $M_{GUT}$ and $\alpha_{GUT}$ as
function of $\tm$ and for this reason in the right panel we do not
differentiate between both cases.  Using the experimental value for
the strong coupling one can get for the case where all gauginos are at
the EW scale the $1\,\sigma$ prediction for $\tm $ as
\be
\tm\simeq 10^{1.6\pm 0.6}\ {\rm TeV} ~ .
\label{pred1}
\ee
If one considers instead the standard unification
relation between the gaugino masses the predicted values of
$\alpha_3(M_Z)$ would be shifted to larger values, $\Delta
\alpha_3(M_Z) \simeq 0.005$ and the resulting values of $\tm$ will be
shifted up to
\be
\tm\simeq 10^{3.3\pm 0.6}\ {\rm TeV} ~ .
\label{pred2}
\ee
For these two ranges of $\tm$ altogether the unification scale
$M_{GUT}$ turns out to be all in all $1\div2 \times 10^{16}$ GeV,
where the smaller value is referred to the largest available $\tm$
value.

The numerical results may be analytically understood by considering the
modifications of the two-loop predictions for $\alpha_3(M_Z)$ by
one-loop threshold corrections induced by the supersymmetric
particles,
\begin{equation}
\left.\left.  \alpha_3(M_Z) = \alpha_3(M_Z)\right|_{\rm MSSM} -
\alpha_3^2(M_Z)\right|_{\rm MSSM} \frac{19}{28 \pi} \ln \left(
\frac{T_{\rm MSSM}}{M_Z} \right),
\end{equation}
where~\cite{CPW}
\begin{equation}
T_{MSSM} = |\mu| \left( \frac{M_2}{M_3} \right)^{28/19}
                 \left(\frac{M_2}{|\mu|} \right)^{4/19}
                 \left(\frac{\tm}{m_{\tilde{t}}}\right)^{5/19}
                 \left(\frac{\tm}{|\mu|}\right)^{3/19}.
\label{TMSSM}
\end{equation}
In the above $\alpha_3(M_Z)|_{\rm MSSM} \simeq 0.127$ would be the
value that would be obtained if all supersymmetric particles would
have masses equal to $M_Z$. The second-to-last and last terms in
Eq.~(\ref{TMSSM}) represent the effects of separating the stop mass
with respect to the other sfermion masses and of increasing the non-SM
Higgs doublet mass, respectively.  In particular for the case in which
all gaugino masses, $|\mu|$ and the light stop are of the order of the
weak scale one can reproduce the $1\sigma$ prediction for $\tm $ as
given in Eq.~(\ref{pred1}), while in the case of standard unification
relation of gaugino masses Eq.~(\ref{pred2}) is recovered if equal
values of $|\mu|$ and $M_2$ of order of the weak scale are assumed.
These low energy supersymmetric threshold effects may be compensated
by thresholds at the GUT scale, which are strongly model dependent,
but are naturally of the same order as the low energy supersymmetric
thresholds (see, for instance Ref.~\cite{Langacker}).  Therefore for
hierarchical gaugino masses, as the ones required by electroweak
baryogenesis, the natural values of $\tm$ necessary to achieve
unification and consistent at 95 \% C.L. with present experimental
values of $\alpha_3(M_Z)$ are about 100~TeV. Somewhat lower values of
$\tm$ of the order of a few tens of TeV may be obtained by pushing
$|\mu|$ to larger values.

The values of $\tm$ consistent at 95 \% C.L. with unification of
couplings (for light gluinos $3\,{\rm TeV}<\tm<600\,{\rm TeV}$ and for
standard gaugino unification $100\,{\rm TeV}<\tm<3\times10^4\,{\rm
TeV}$) have an impact on the Higgs mass predictions. From
Fig.~\ref{plotStH}, we can see that values of $\tm$ larger than a few
TeV lead to consistency with the LEP Higgs mass constraints for a
large range of values of $\tan\beta$ when $A_t/\tm$ is conveniently
chosen.  For the (positive) value of $M_U^2$ used in
Fig.~\ref{plotStH}, $M_U^2= (200~ {\rm GeV})^2$, values of $\tilde m$
of the order of 600 TeV ($3\times10^4$ TeV) lead to maximum values of
the Higgs mass around 144 (150) GeV.  On the other hand, as shown in
Fig.~\ref{plot100}, for negative values of $M_U^2$ and $A_t\lesssim
0.5~\tm$, as required by electroweak baryogenesis, the values
$\tm\simeq 600$ TeV ($3\times 10^4$ TeV) lead to maximum values of
the Higgs boson mass around $125 ~(129)$~GeV.

\section{\sc Model cosmology and collider phenomenology}
\label{pheno}

The cosmology and collider phenomenology of the light stop scenario
has been the subject of study of different articles.  For masses below
135~GeV, as preferred by the electroweak baryogenesis scenario, the
light stop mass will be in general smaller than the sum of the
$W$~mass, the $b$-quark mass and the lightest neutralino mass, and
therefore its three body decay channels will be suppressed. Under
these conditions, the main stop decay channel may be a loop-induced two
body decay channel into a charm quark and the lightest neutralino.
Searches for such a light stop at LEP put a bound on its mass of about
100~GeV~\cite{Kraan:2003it}.

Current searches at the Tevatron collider for a stop decaying into
charm jets and neutralinos lead to a final state of two jets and
missing energy. The jets should be sufficiently energetic for the
Tevatron to be able to trigger on those events, what in practice
demands mass differences between the stops and the neutralinos of
about 30 GeV or larger~\cite{Aaltonen:2007sw,Abazov:2008rc}.
Therefore the Tevatron collider cannot set any constraints on direct
production of stops for mass differences smaller than 30~GeV. Searches
for light stops in direct pair production of these particles, will be
equally difficult at the LHC.

Small mass differences between the stop and the neutralino define a
particularly interesting region of parameters since they are helpful
in providing the proper dark matter density in scenarios with heavy
fermions. Indeed, for mass differences of about 20~GeV, the
co-annihilation between the stop and the lightest neutralino leads to
a neutralino dark matter density consistent with experimental
observations~\cite{Balazs}.

Searches for light stops at the LHC may proceed through additional
production channels. For instance, the light stops may be produced
from the decay of heavy gluinos. Assuming that the right-handed stops
are the only squarks with masses below the gluino mass, as happens in
the light stop scenario discussed in this article, the gluinos being
Majorana particles may decay into a stop and an anti-top or into an
anti-stop and a top-quark. One can then consider the decay of a pair
of gluinos into two equal sign top-quarks and two stops (two charm
jets and missing energy). It has been shown~\cite{Sabine} that under
these conditions, the light stops may be found even for small mass
differences, of about 5~GeV, provided the heavy gluinos are lighter
than about 900~GeV.

One would be interested in finding a method of stop detection that
would be independent of the exact masses of other sparticles and which
would work for small mass differences. A possibility is to analyze the
possible production of light stops in association with a photon or a
gluon (jet). The photon signatures are particularly clean, and for
small mass differences they may be used as a complementary channel for
the search for light stops at hadron colliders leading to a final
state of two photons, soft jets and missing energy. Although not as
clean as the photon signatures, due to larger rates, the jet plus
missing energy signature may allow a further extension of the LHC
reach for light stops.  An analysis in this direction is in
progress~\cite{Ayres}.

\section{\sc Conclusions}
\label{conclusion}

   In this article we analyzed the light stop scenario in which all
squarks and sleptons, apart from a mainly right-handed stop, are
significant heavier than the weak scale. The large values of the
scalars imply that the low energy effective theory predictions may
only be evaluated in a precise way by resummation of the large leading
logarithms associated with the decoupling of the heavy scalars. Since
supersymmetry is broken at scales below the heavy scalar mass $\tm$,
the Yukawa couplings associated with gauginos and Higgsinos must be
computed, starting with their boundary values given by the gauge and
supersymmetric Yukawa couplings respectively. Similarly the quartic
couplings of the Higgs boson and the light top-squark may be computed
through their RG evolution to lower energies.

We have applied the low energy effective theory to obtain a reliable
computation of the lightest CP-even Higgs boson mass for large values
of $\tm$. In the extreme case where $\tm$ is close to the EW scale,
logarithm resummation is unnecessary and we have checked that our
calculation of the SM-like Higgs is consistent with earlier
calculations in the literature~\cite{Carena:1995bx}. Since the quartic
coupling is bounded by its relation with the weak gauge couplings plus
finite threshold corrections at high energies, the Higgs mass remains
bounded to small values, smaller than about 133~GeV for negative
values of $M_U^2 \simeq -(100\; {\rm GeV})^2$ and $A_t\lesssim0.5~\tm$ even
for large values of $\tm$. This has important implications for the
realization of the electroweak baryogenesis scenario. In a general
light stop scenario with no EWBG mechanism built in, this bound on the
Higgs mass may be relaxed for positive values of $M_U^2$, for which
the trilinear mass parameter $A_t$ may be pushed to larger values,
leading to masses that may be as large as 152~GeV for large values of
$\tm$.

We have also analyzed the issue of unification of gauge couplings. We
have shown that the corrections induced by the heavy spectrum are
helpful in rendering the measured values of the gauge couplings
consistent with the unification conditions even for relatively large
values of $\tm$. For instance considering universal values of the
gaugino masses at low energies and values of $|\mu|$ of about 100~GeV,
one obtains that appropriate unification is achieved for values of
$\tm \simeq 10^{1.6 \pm 0.6}$~TeV.  A similar result is obtained for
the standard unification relation between the gaugino masses, for a
Higgsino mass parameter of about 1~TeV.  If $|\mu|$ takes values close
to 100~GeV, instead, the value of $\tm$ consistent with the
unification of couplings is pushed up to $\tm \simeq 10^{3.3 \pm
0.6}$~TeV. This ranges of masses have implications on the predicted
Higgs mass. For positive values for $M_U^2 \simeq (200\,{\rm GeV})^2$,
the ranges of values of $\tm$ compatible at 95 \% C.L. with gauge
coupling unification lead to an upper bound on the Higgs mass of about
$150$~GeV. For negative values of $M_U^2$ and $A_t\lesssim0.5~\tm$, as
the ones required for baryogenesis, the upper bound becomes even
stronger, of about $129$~GeV.

The resulting phenomenology of the light stop scenario was also
discussed in some detail. If a light Higgs, with mass
$m_h\lesssim 133$~GeV is found, the next step to confirm the EWBG
scenario within the MSSM would be the discovery of a light stop, with
a mass below the top quark mass. Light stop searches at the Tevatron
may lead to an experimental confirmation of this scenario, but may not
be successful if the mass difference between the stop and the
neutralino is smaller than about 30~GeV.  Unfortunately, these small
mass differences may be the ones required to obtain the proper dark
matter relic density by means of coannihilation between the stop and
the lightest neutralino. Searches at the LHC may be able to test the
coannihilation region in case the gluino is lighter than about
900~GeV. For heavier gluino masses alternative methods of detection at
the LHC via the production of light stops in association with photons
or jets, are currently being studied and seem to be promising.

\subsection*{\sc Acknowledgments}
\noindent 
We would like to thank C.~Balazs, A.~Freitas, T.~Konstandin,
D.~Morrisey and A.~Menon for useful discussions and interesting
observations.  Work supported in part by the European Commission under
the European Union through the Marie Curie Research and Training
Networks ``Quest for Unification" (MRTN-CT-2004-503369) and
``UniverseNet" (MR-TN-CT-2006-035863). Work at ANL is supported in
part by US DOE, Division of HEP, Contract DE-AC-02-06CH11357 and
Fermilab is operated by Fermi Research Alliance, LLC under Contract
No.  DE-AC02-07CH11359 with the United States Department of Energy.
The work of M.Q. was partly supported by CICYT, Spain, under contract
FPA 2005-02211.

\section*{\sc Appendix}

\appendix
\section{\sc Thresholds}
\label{thresholds}
In this appendix we will give some details about the calculation of
the different thresholds which appear in Section~\ref{effective}.  At
the renormalization scale $\tm$ the ET (\ref{lagreff}) has to describe
the same physics as its corresponding HE theory, the MSSM. Here we
will match both theories at one--loop in the Landau gauge using the
step--function approximation.  Graphically the matching is presented
in Figs.~\ref{figQ}-\ref{figK}.

Since the matching has to be performed order by order in perturbation
theory, we will start by considering the effective coupling $Q$, which
is the only one having not trivial matching at tree--level (see
Fig.~\ref{figQ}). This matching can be easily solved by neglecting
heavy field kinetic--terms and solving for their equation of
motion. The result is given in (\ref{deltaQ}).

At one--loop we cannot follow the same procedure because in
dimensional regularization no kinetic--term can be neglected inside
the loop. In such a case we can compute the diagrams in the LE and HE
theories and, after using the tree--level matching conditions, impose
the equivalence of the two results~\footnote{Clearly, this operation
is well defined only after fixing the subtraction scheme, in our case
the $\overline{MS}$.}.  We will follow this diagrammatic approach only
for the $h_t,~Y_t,~G$ proper vertex matching,
Figs.~\ref{fight}-\ref{figG}, and the wave function contribution
furnished by each external leg (as an example we draw the case of
$\str$ and $t_R$ in Figs.~\ref{figst} and \ref{figt}). In fact in
these cases identifying the threshold sources is straightforward. The
corresponding HE diagrams are shown in
Figs.~\ref{fight}-\ref{figt}. The resulting proper vertex thresholds
are given in (\ref{deltaht})-(\ref{deltaY}) and the wave function ones
in (\ref{wfgl}).  Finally for the proper vertex threshold
$\Delta\lambda$ ($\Delta K$) it is easier to match the one--loop Higgs
(stop) LE and HE effective potentials, instead of performing the
matching diagrammatically (see Fig.~\ref{figK})~\footnote{In the $Q$
matching condition we do not consider the one--loop proper vertex
threshold since the tree--level one dominates.}.

Let us start to explicitly analyze the case of $\Delta\lambda$ for
which we have to impose the equivalence of the terms proportional to
$\phi_c^4$ after the expansion of the HE and LE effective potentials
\bea
\label{LamPots}
\Lambda-\frac{m^2}{2}\phi_c^2+\frac{\lambda}{8}\phi_c^4 + V_{LE}^h(\phi_c) =
\Lambda'-\frac{m'^2}{2}\phi_c^2+ 
\frac{\left(g^2+g^{\prime 2}\right)}{32} \phi_c^4
\cos^2 2\beta +   V_{HE}^h(\phi_c) +{\cal O}(\phi_c^5) \nonumber\\
\eea
where $m^2$ and $\Lambda$ are equal to $m'^2 $ and $\Lambda'$ up to 
threshold effects coming from the difference between the one-loop
contributions $V_{HE}^h(\phi_c)$ and $V_{LE}^h(\phi_c)$
\bea
\label{diffPot}
&& V_{HE}^h(\phi_c)-V_{LE}^h(\phi_c)\nonumber\\
&&= 
\frac{6}{64 \pi^2} \sum_{r=\tilde t_1,\tilde t_2}
           m_{r}^4 \left(
                   \ln \frac{m_{r}^2}{\tau^2}
                             -\frac{3}{2}\right) -
\frac{6}{64 \pi^2}
           ~m_{\tilde t_R}^4 \left(
                   \ln \frac{m_{\tilde t_R}^2}{\tau^2}
                             -\frac{3}{2}\right)\ ,
\eea
where the renormalization scale is fixed to $\tau=\tm$, $m_{\str}$ is
explicit in (\ref{stopmass}) and $\tilde t_1,\tilde t_2$ are the
eigenvalues of
\bea
\label{m2st}  \left(\begin{array}{cc} 
    \tm^2 + \lambda_t^2 (\phi_c^2/2)  \sin^2\beta &
    -\lambda_t \tilde A_t (\phi_c/\sqrt2)  \sin\beta\\ \\
    -\lambda_t \tilde A_t  (\phi_c/\sqrt2)  \sin\beta  &
     M_U^2 + \lambda_t^2 (\phi_c^2/2)  \sin^2\beta
  \end{array} \right)   ~,
\eea
with $\tilde A_t=A_t-\mu/\tbe$.  The threshold $\Delta\lambda$ can be
derived extracting the coefficient of the term $\phi_c^4/8$ from the
right-hand side of (\ref{diffPot}). Finally remembering that $\tm^2
\gg M_U^2$, we obtain the relation (\ref{deltalam}).

\begin{figure}
\begin{center}

\vspace{-10pt}  \hfill \\

\SetScale{0.3}
\begin{picture}(70,0)(0,27)
\DashLine(0,200)(200,0){13}\Text(-5,70)[]{$H$} \Text(70,70)[]{$H$}
\DashLine(0,0)(200,200){13}\Text(-5,-8)[]{$\tilde t_R$} 
\Text(70,-8)[]{$\tilde t_R$}
\Vertex(100,100){8}
\end{picture} $+~~~\cdots~~~ = ~~~$
\begin{picture}(70,30)(0,27)
\DashLine(0,200)(200,0){13}\Text(-5,70)[]{$H$} \Text(70,70)[]{$H$}
\DashLine(0,0)(200,200){13}\Text(-5,-8)[]{$\tilde t_R$} 
\Text(70,-8)[]{$\tilde t_R$}
\end{picture} \vspace{10pt}$~~+~~~~$\vspace{-10pt}
\begin{picture}(70,30)(0,27)
\DashLine(0,200)(100,100){13}\Text(-5,70)[]{$H$} \Text(100,70)[]{$H$}
\DashLine(0,0)(100,100){13}\Text(-5,-8)[]{$\tilde t_R$} 
\Text(100,-8)[]{$\tilde t_R$}
\DashLine(100,100)(200,100){13} \Text(46,40)[]{$\tilde q$}
\DashLine(200,100)(300,200){13}
\DashLine(200,100)(300,0){13} 

\end{picture} $~~~~~~~~~+ ~~~\cdots$
\SetScale{1}
\vspace{26pt}  \hfill \\
\end{center}
\caption{Tree--level proper vertex matching of $Q$. \label{figQ}}

\begin{center}
\vspace{40pt}  \hfill \\
\SetScale{0.3}

\begin{picture}(70,0)(0,27)
\Line(200,0)(100,90)\Text(70,-8)[]{$t_R$}
\Line(0,0)(100,90)\Text(-5,-8)[]{$t_L$} 
\DashLine(100,90)(100,200){13} \Text(30,70)[]{$H$} 
\Vertex(100,90){8}
\end{picture}
 $~+ ~~~\cdots ~~~~ =~~~$
\begin{picture}(70,0)(0,27)
\Line(200,0)(100,90)\Text(70,-8)[]{$t_R$}
\Line(0,0)(100,90)\Text(-5,-8)[]{$t_L$} 
\DashLine(100,90)(100,200){13} \Text(30,70)[]{$H$}
\end{picture}
$~+ ~~~~~$
\begin{picture}(70,0)(0,27)
\Line(200,0)(200,40)\Text(70,-8)[]{$t_R$}
\Line(0,0)(0,40) \Text(-5,-8)[]{$t_L$} 
\DashLine(100,150)(100,200){10} \Text(30,70)[]{$H$}

\Photon(200,40)(0,40){13}{6}
\Line(200,40)(0,40) \Text(31,3)[]{$\tilde g$}
\DashLine(0,40)(100,150){13}\Text(53,33)[]{$\tilde t_R$}
\DashLine(200,40)(100,150){13} \Text(8,33)[]{$\tilde t_L$}
\end{picture}
$~+ ~~~\cdots$
\SetScale{1}
\vspace{26pt}  \hfill \\
\end{center}
\caption{One--loop proper vertex matching of $h_t$ at $\tau=\tm$.\label{fight}}

\begin{center}
\vspace{40pt}  \hfill \\
\SetScale{0.3}

\begin{picture}(70,0)(0,27)
\Line(200,0)(100,90)\Text(70,-8)[]{$t_L$}
\Line(0,0)(100,90)\Text(-5,-8)[]{$\tilde H_u$} 
\DashLine(100,90)(100,200){13} \Text(30,70)[]{$\tilde t_R$} 
\Vertex(100,90){8}
\end{picture}
 $~+ ~~~\cdots ~~~~ =~~~$
\begin{picture}(70,0)(0,27)
\Line(200,0)(100,90)\Text(70,-8)[]{$t_L$}
\Line(0,0)(100,90)\Text(-5,-8)[]{$\tilde H_u$} 
\DashLine(100,90)(100,200){13} \Text(30,70)[]{$\tilde t_R$}
\end{picture}
$~+ ~~~~~$
\begin{picture}(70,0)(0,27)
\Line(200,0)(200,40)\Text(70,-8)[]{$t_L$}
\Line(0,0)(0,40) \Text(-5,-8)[]{$\tilde H_u$} 
\DashLine(100,150)(100,200){10} \Text(30,70)[]{$\tilde t_R$}

\DashLine(200,40)(0,40){13} \Text(31,3)[]{$\tilde t_L$}
\Photon(200,40)(100,150){13}{6}
\Line(0,40)(100,150) \Text(8,33)[]{$t_R$}
\Line(200,40)(100,150) \Text(54,33)[]{$\tilde g$}
\end{picture}
$~+ ~~~\cdots$
\SetScale{1}
\vspace{26pt}  \hfill \\
\end{center}
\caption{One-loop proper vertex matching of $Y_t$ at $\tau=\tm$.\label{figYt}}

\begin{center}
\vspace{40pt}  \hfill \\
\SetScale{0.3}

\begin{picture}(70,0)(0,27)
\Line(200,0)(100,90)\Text(70,-8)[]{$t_R$}
\Photon(0,0)(100,90){13}{6}
\Line(0,0)(100,90)\Text(-5,-8)[]{$\tilde g$} 
\DashLine(100,90)(100,200){13} \Text(30,70)[]{$\tilde t_R$} 
\Vertex(100,90){8}
\end{picture}
 $~+ ~~~\cdots ~~~~ =~~~$
\begin{picture}(70,0)(0,27)
\Line(200,0)(100,90)\Text(70,-8)[]{$t_R$}
\Photon(0,0)(100,90){13}{6}
\Line(0,0)(100,90)\Text(-5,-8)[]{$\tilde g$} 
\DashLine(100,90)(100,200){13} \Text(30,70)[]{$\tilde t_R$} 
\end{picture}
$~+ ~~~~~$
\begin{picture}(70,0)(0,27)
\Line(200,0)(200,40)\Text(70,-8)[]{$t_R$}
\Photon(0,0)(0,40){13}{2.5}
\Line(0,0)(0,40) \Text(-5,-8)[]{$\tilde g$} 
\DashLine(100,150)(100,200){10} \Text(30,70)[]{$\tilde t_R$}

\DashLine(200,40)(0,40){13} \Text(31,3)[]{$\tilde t_L$}
\Line(0,40)(100,150) \Text(8,33)[]{$t_L$}
\Line(200,40)(100,150) \Text(54,33)[]{$\tilde H_u$}
\end{picture}
$~+ ~~~\cdots$
\SetScale{1}
\vspace{26pt}  \hfill \\
\end{center}
\caption{One--loop proper vertex matching of $G$ at $\tau=\tm$.\label{figG}}

\end{figure}

\begin{figure}[h!]

\begin{center}

\vspace{10pt}  \hfill \\

\SetScale{0.3}

\hspace{-4mm}
\begin{picture}(70,0)(0,27)
\DashLine(0,100)(200,100){13}\Text(33,39)[]{$\tilde t_R$} 
\Vertex(100,100){8}
\end{picture} 
$+~~\cdots~~=~$
\begin{picture}(70,0)(0,27)
\DashLine(0,100)(170,100){13}\Text(26,39)[]{$\tilde t_R$} 
\end{picture} 
\hspace{-4mm}$+~~~$
\begin{picture}(70,30)(0,27)
\DashLine(0,100)(60,100){13}\Text(-2,39)[]{$\tilde t_R$}
\DashCArc(110,100)(50,0,180){13}\Text(35,54)[]{$\tilde q_L$}
\DashCArc(110,100)(50,180,0){13}\Text(35,5)[]{$H_h$}
\DashLine(160,100)(220,100){13}\Text(70,39)[]{$\tilde t_R$}
\end{picture} 
$~~+ ~~$
\begin{picture}(70,30)(0,27)
\DashLine(0,100)(60,100){13}\Text(-2,39)[]{$\tilde t_R$}
\DashCArc(110,100)(50,0,180){13}\Text(35,54)[]{$\tilde q_L$}
\DashCArc(110,100)(50,180,0){13}\Text(35,5)[]{$H$}
\DashLine(160,100)(220,100){13}\Text(70,39)[]{$\tilde t_R$}
\end{picture} 
$~~+ ~\cdots$
\SetScale{1}
\vspace{10pt}  \hfill \\
\end{center}
\caption{One-loop matching of the $\str$ wave function renormalization
  at $\tau=\tm$.
\label{figst}}

\begin{center}

\vspace{10pt}  \hfill \\

\SetScale{0.3}

\hspace{-4mm}
\begin{picture}(70,0)(0,27)
\Line(0,100)(200,100)\Text(33,39)[]{$t_R$} 
\Vertex(100,100){8}
\end{picture} 
$+~~\cdots~~=~$
\begin{picture}(70,0)(0,27)
\Line(0,100)(170,100)\Text(16,39)[]{$t_R$} 
\end{picture} 
\hspace{-4mm}$+~~~$
\begin{picture}(70,30)(0,27)
\Line(0,100)(60,100)\Text(-2,39)[]{$t_R$}
\CArc(110,100)(50,0,180)\Text(35,51)[]{$q_L$}
\DashCArc(110,100)(50,180,0){13}\Text(35,5)[]{$H_h$}
\Line(160,100)(220,100)\Text(70,39)[]{$t_R$}
\end{picture} 
$~~+ ~~$
\begin{picture}(70,30)(0,27)
\Line(0,100)(60,100)\Text(-2,39)[]{$t_R$}
\CArc(110,100)(50,0,180)\Text(35,54)[]{$\tilde H_u$}
\DashCArc(110,100)(50,180,0){13}\Text(35,5)[]{$\tilde q_L$}
\Line(160,100)(220,100)\Text(70,39)[]{$t_R$}
\end{picture} 
$~~+ ~\cdots$
\SetScale{1}
\vspace{10pt}  \hfill \\
\end{center}
\caption{One-loop matching of the $t_R$ wave function renormalization at $\tau=\tm$.
\label{figt}}
\vspace{1.3cm}

\begin{center}

\vspace{0pt}  \hfill \\

\SetScale{0.3}

\begin{picture}(70,0)(0,27)
\DashLine(0,200)(200,0){13}\Text(-5,70)[]{$H(\tilde t_R)$} 
\Text(70,70)[]{$H(\tilde t_R)$}
\DashLine(0,0)(200,200){13}\Text(-5,-8)[]{$H(\tilde t_R)$} 
\Text(70,-8)[]{$H(\tilde t_R)$}
\Vertex(100,100){8}
\end{picture} $~~~~~+~~~~~~~$
\begin{picture}(70,30)(0,27)
\DashLine(0,200)(100,100){13}\Text(-5,70)[]{$H(\tilde t_R)$} 
\Text(100,70)[]{$H(\tilde t_R)$}
\DashLine(0,0)(100,100){13}\Text(-5,-8)[]{$H(\tilde t_R)$} 
\Text(100,-8)[]{$H(\tilde t_R)$}
\DashCArc(150,100)(50,0,360){13} \Text(46,55)[]{$\tilde t_R(H)$} 
\Text(46,5)[]{$\tilde t_R(H)$}
\DashLine(200,100)(300,200){13}
\DashLine(200,100)(300,0){13}
\Vertex(100,100){8}
\Vertex(200,100){8} 

\end{picture} $~~~~~~~~~+ ~~~\cdots ~~~~ =~~~~$
\begin{picture}(70,0)(0,27)
\DashLine(0,200)(200,0){13}\Text(-5,70)[]{$H(\tilde t_R)$} 
\Text(70,70)[]{$H(\tilde t_R)$}
\DashLine(0,0)(200,200){13}\Text(-5,-8)[]{$H(\tilde t_R)$} 
\Text(70,-8)[]{$H(\tilde t_R)$}
\end{picture} 
\vspace{80pt} \\
${}\hspace{80pt} + ~~~$
\begin{picture}(70,30)(0,27)
\DashLine(0,200)(100,100){13}\Text(-5,70)[]{$H(\tilde t_R)$} 
\Text(100,70)[]{$H(\tilde t_R)$}
\DashLine(0,0)(100,100){13}\Text(-5,-8)[]{$H(\tilde t_R)$} 
\Text(100,-8)[]{$H(\tilde t_R)$}
\DashCArc(150,100)(50,0,360){13} \Text(46,55)[]{$\tilde t_R(H)$} 
\Text(46,5)[]{$\tilde t_R(H)$}
\DashLine(200,100)(300,200){13}
\DashLine(200,100)(300,0){13}
\end{picture}
$~~~~~~~~~~+~~~~~~~$
\begin{picture}(70,30)(0,27)
\DashLine(0,200)(50,150){13}\Text(-5,70)[]{$H(\tilde t_R)$} 
\Text(70,70)[]{$H(\tilde t_R)$}
\DashLine(0,0)(50,50){13}\Text(-5,-8)[]{$H(\tilde t_R)$} 
\Text(70,-8)[]{$H(\tilde t_R)$}
\DashLine(50,50)(50,150){13}\Text(-1,30)[]{$\tilde t_R(H)$}
\DashLine(50,50)(150,50){13}\Text(65,30)[]{$\tilde t_R(H)$}
\DashLine(50,150)(150,150){13}\Text(30,5)[]{$\tilde q$}
\DashLine(150,50)(150,150){13}\Text(30,55)[]{$\tilde q$}
\DashLine(150,150)(200,200){13}
\DashLine(150,50)(200,0){13}
\end{picture}
$~~~~~~+ ~~~\cdots$
\SetScale{1}
\vspace{35pt}  \hfill \\
\end{center}
\caption{One-loop proper vertex matching of $\lambda$ ($K$) at $\tau=\tm$. \label{figK}}

\end{figure}

Following the same idea we can also obtain $\Delta K$. We give a
constant background $s_c$ to the real third colour component of
$\tilde t_R$, {\it i.e.} $\langle\tilde t_{R_3}\rangle=s_c/\sqrt{2}$,
which breaks the $SU(3)_c$ and $U(1)_Y$ symmetries, and we impose the
equivalence of its one--loop effective potential in the LE and HE
theory at the scale $\tm$
\bea
\label{Kpots}
\Lambda+\frac{M_U^2}{2}s_c^2 + \frac{K}{24} s_c^4 + V_{LE}^\str(s_c) 
  =  \Lambda'+\frac{{M'}_U^2}{2}s_c^2 +\frac{g_3^2}{24} s_c^4 +
 V_{HE}^\str(s_c) + {\cal O}(s_c^5) ~~,
\eea
where 
\bea
\label{diffpotK}
V^\str_{HE}-V^\str_{LE} =
 \frac{4}{64 \pi^2} \sum_{r=1}^5
           m_{r}^4 \left(
                   \log \frac{m_{r}^2}{\tau^2}
                             -\frac{3}{2}\right) - 
\frac{4}{64 \pi^2}
           ~\nu_H^4 \left(
                   \log \frac{\nu_H^2}{\tau^2}
                             -\frac{3}{2}\right)  ~,
\eea
with $\tau=\tm$ and $\nu^2_{H}=\lambda_t^2 \sin^2\beta (1-\frac{\tilde
A_t^2}{\tm^2}) \frac{s_c^2}{2}$. Moreover for $r=1,2,3$~ the masses
$m_r^2$ are the eigenvalues of the squared mass matrix of $\tilde
q_3$, $H$ and $H_h$ (the heavy projection of the Higgses:
$H_u^\dagger\rightarrow \sbe H_h^t \epsilon $ and $H_d\rightarrow \cbe
\epsilon H_h^*$)
\bea
{\cal N}&=& \left(
            \begin{array}{ccc}
	      \left(\lambda_t^2-\frac{g^2_3}{3}\right) \frac{s_c^2}{2}+\tm^2
	      &
              \lambda_t \tilde B_t  \frac{s_c}{\sqrt{2}}\cbe
	      &
	      \lambda_t \tilde A_t  \frac{s_c}{\sqrt{2}}\sbe  \\
	      \lambda_t \tilde B_t  \frac{s_c}{\sqrt{2}}\cbe
              &
	      \tm^2 +\lambda_t^2 \frac{s_c^2}{2}\cos^2\beta
	      &
	      \lambda_t^2  \frac{s_c^2}{4} \sin2\beta\\
	      \lambda_t \tilde A_t  \frac{s_c}{\sqrt{2}}\sbe
              &
	      \lambda_t^2  \frac{s_c^2}{4} \sin2\beta
	      &
	      \lambda_t^2 \frac{s_c^2}{2} \sin^2\beta
	    \end{array}
	    \right) ~,
\eea
with $\tilde B_t=A_t+\mu \tbe$, and finally $m_4^2\equiv m_{\tilde
q_1}^2=m_5^2\equiv m_{\tilde q_2}^2=\tm^2+ \frac{s_c^2}{12}$.
Extracting from the right-hand side of (\ref{diffpotK}) the
coefficient of the term $s_c^4/24$ we obtain $\Delta K$ as expressed
in (\ref{deltaK}).

To conclude here we collect all the proper vertex thresholds 
\be
\label{deltaQ}
\Delta Q=-  \lambda_t^2 \sin^2\beta \frac{\left|\tilde A_t\right|^2}{\tm^2}
~,
\ee
\be
\Delta h_t=\frac{8}{3 (4 \pi)^2} g_3^2 \lambda_t\tilde A_t\sbe
~M_3\,b_1 ~,
\label{deltaht}
\ee
\be
\Delta G = \frac{2}{(4\pi)^2} g_3 \lambda_t^2 \left(-1+ 
       \frac{\mu^2 \ln(\mu^2/\tm^2)}{\mu^2-\tm^2}
       \right) ~,
\label{deltaG}
\ee
\be
\Delta Y_t = \frac{8}{3(4\pi)^2} g_3^2 \lambda_t \left(1- 
       \frac{M_3^2 \ln(\tm^2/M_3^2)}{\tm^2-M_3^2}\right)
\label{deltaY} ~,
\ee
\be
\label{deltalam}
\Delta\lambda= \frac{3}{8\pi^2} (\lambda_t \sbe)^4 \tilde A_t^4~
\frac{\tm^2-M_U^2 - M_U^2 \ln(\tm^2/M_U^2)}{(\tm^2-M_U^2)^3}~,
\ee
\be
\label{deltaK}
\Delta K = c_0 +c_1 \frac{\tilde A_t}{\tm} + c_2 \frac{\tilde
A_t^2}{\tm^2} + c_3 \frac{\tilde A_t^4}{\tm^4} ~~,
\ee
where
\bea
b_1&=&\frac{M_3^2
  (\tm^2-M_U^2)\log(\tm^2/M_3^2)+M_U^2(\tm^2-M_3^2)
\log(\tm^2/M_U^2)}{(\tm^2-M_3^2)(M_3^2-M_U^2)(\tm^2-M_U^2)} ~,
\nonumber\\
c_0&=&\frac{1}{16\pi^2}\left({3 \lambda_t^4 \sin^2 2\beta}+{2 \lambda_t^2 \cos ^2\beta
\left[g_3^2-3 \lambda_t^2\left(1+ \cos^2\beta\right)\right]
\frac{\tilde B_t^2}{\tm^2}}+{\lambda_t^4
\cos^4\beta~ \frac{\tilde B_t^4}{\tm^4}}\right) ~,
\nonumber\\
c_1&=&-\frac{3 \lambda_t^4 \sin^2 2\beta}{8 \pi ^2}\frac{\tilde B_t}{\tm} ~,
\nonumber\\
c_2&=&\frac{1}{32\pi^2}\left({8 \lambda_t^2 \sin^2 \beta \left[g_3^2-3 \lambda_t^2\left(1-
\sin^2\beta\right)\right]}+{3 \lambda_t^4 \sin ^2 2\beta~
\frac{\tilde B_t^2}{\tm^2}}\right) ~,
\nonumber\\
c_3&=&-\frac{3 \lambda_t^4 \sin ^4\beta}{4\pi^2} 
\nonumber ~,
\eea
along with the wave function threshold contributions of each external
leg
\bea
\begin{array}{lr}
Z_{\tilde t_R}=2 \left(\lambda_t \tilde B_t\cbe\right)^2 F(\tm^2)
               +2 \left(\lambda_t \tilde A_t\cbe\right)^2 F(0) 
&\qquad [\tilde qH_h+\tilde qH],
\\
Z_H= \left(\lambda_t \tilde A_t\cbe\right)^2 F(M_U^2)
&[\tilde q ~\tilde t_R],
\\
Z_{t_R}= 2 \left(\lambda_t \cbe\right)^2 E(0) +2 \lambda_t^2~ E(\mu^2) 
&[ H_h q+\tilde q\tilde H_u],
\\
Z_{t_L}=  \left(\lambda_t \cbe\right)^2 E(0) +\frac{8}{3}g_3^2~ E(M_3^2)
& [H_h t_R+\tilde q \tilde g],\\
Z_{\tilde H_u}=\lambda_t^2~ E(0) 
&[\tilde q t_R],\\
Z_{\tilde g}=11  g_3^2~ E(0) 
&[\tilde q q],
\end{array}
\label{wfgl}
\eea
where the particles propagating in the loops are indicated inside
squared brackets and the functions $F(m^2)$ and $E(m^2)$ are defined
by
\bea
F(m^2)&\equiv& - \frac{1}{\tm^2 (4\pi)^2} 
     \frac{a^4-1-2 a^2 \log(a^2)}{2(a^2-1)^3} ~,\\
E(m^2)&\equiv& -\frac{1}{(4\pi)^2}
\frac{-1+(4-3a^2)a^2+2 a^2 \log a^2}{4 (a^2-1)^2}~, 
\eea
with $a^2=m^2/\tm^2$.

Because of the thresholds (\ref{wfgl}) the kinetic terms of the
effective theory would not be canonically normalized if these wave
function thresholds were not absorbed in a redefinition of the
effective fields. This implies that any generic effective coupling
$\rho$ has also gotten a wave function threshold dependence coming from
its field redefinitions as
\bea
 1-\frac{1}{2}\Delta Z_\rho \equiv 1-\frac{1}{2}\sum_i Z_i  ~~, 
\eea
where $i$ runs over the fields of the interaction $\rho$.

A last remark concerns the gauge couplings. They have no threshold
because Ward identities impose a cancellation between the proper
vertex threshold and the non-vector fields wave function
ones. Therefore a possible threshold could only come from the vector
boson wave function threshold but the latter is zero when evaluated at
the renormalization scale $\tm$.  Finally let us observe that the mass
thresholds are not necessary for our aim. In fact the LE masses only
appear inside one--loop thresholds in which a possible one--loop mass
thresholds would only contribute at two--loop.

Finally if we assume the gluino mass heavy enough (but below $\tm$),
it is necessary to also integrate it out and repeat at the scale
$\tau=M_3$ the procedure just described. The gluino decoupling affects
the proper vertex $K$ and the right--handed top and stop propagators,
which produce the right--handed top and stop wave function thresholds
and the mass threshold $\Delta'M_U^2$~\footnote{We also consider the
mass thresholds because we want to know the masses evolution beyond
the decoupling scale $M_3$.}.

Concerning the wave function thresholds, the matching conditions at
$\tau=M_3$ lead to
\bea
\Delta'Z_{t_R} &=& \frac{5 ~G^2}{6 (4\pi)^2} ~, \\
\Delta'Z_{\str} &=& \frac{2 ~G^2}{3 (4\pi)^2} \frac{3-4 b^2+b^4-2
  b^2(b^2-2)\log b^2}{(b^2-1)^2} ~, \nonumber
\eea
where $b^2=M_U^2/M_3^2$.

In order to calculate the proper vertex threshold $\Delta'K$ and
$\Delta'M_U^2$ we use the procedure of matching the stop effective
potential in the presence of a background field. After giving a VEV to the
third colour stop, $\langle\tilde t_{R_3}\rangle=s_c/\sqrt{2}$, mixing
mass terms between right top and gluino are generated but, after
diagonalizing, only $t_R^{(3)}$ and $\tilde g^{(8)}$ have masses
depending on $s_c$; explicitly $r_{\pm}=M_3\pm\sqrt{G^2 s_c^2\, 4/3
+M_3^2}$~. Therefore the thresholds can come only from the
contribution to the effective potential of the heaviest fermionic
eigenstate, which results
\bea
 - \frac{2 r_+^4}{64 \pi^2}\left[\log\frac{r_+^2}{M_3^2}
   -\frac{3}{2}\right]= 
       \frac{3 ~M_3^4}{64 \pi^2} 
       + \frac{G^2 ~M_3^2}{24\pi^2}s_c^2
       - \frac{G^4}{144\pi^2}s_c^4 + {\cal O}(s_c^6) ~,
\eea
and thus
\bea
\label{thresholdM3}
\Delta' K &=& -\frac{G^4}{6 \pi^2} ~, \nonumber \\
\Delta' M_3 &=& \frac{G^2 ~M_3^2}{24\pi^2} ~.
\eea

\section{\sc Renormalization group equations}
\label{RGE}

In this appendix we sketch the calculation of the one--loop RGE in the
ET~\footnote{We have checked that our results are consistent with the
MSSM~\cite{Martin:1993zk} and Split
Supersymmetry~\cite{ArkaniHamed:2004fb} limits.}.  In order to present
our result it is useful to define
\bea
\label{beta}
	\beta_\eta \equiv \frac{\partial \eta}{\partial \ln \tau}=
	\beta_\eta^{(v)} + \eta \sum_i \frac{n_i}{2} \gamma_{\rho_i}
	~~,
\eea
where $\tau$ is the renormalization scale, $\eta$ is the coupling
between different fields $\rho_i$ with multiplicity $n_i$ where the
index $i$ runs over the fields which are involved in the particular
vertex, and the functions $\beta_\eta^{(v)}$ and $\eta
\gamma_{\rho_i}$ are the respective contributions from the
renormalization of the proper vertex and the anomalous dimension of
each external leg.  In the same way as for the threshold effects the
$\beta_{\eta}$ and $\gamma_{\rho_i}$ functions are computed in the
$\overline {MS}$ renormalization scheme and using the Landau gauge.
We will implicitly consider renormalization scales $\tau$ larger than
any fermionic mass, in particular the gluino mass $M_3$, and for
$\tau<M_3$ the correct results are obtained by simply erasing the
couplings $G$ and $M_3$ and disregarding $\beta_{\tilde g}$ and
$\gamma_{\tilde g}$.

By using the diagrammatic procedure we find
\bea
\label{betaF}
	&&(4\pi)^2 \gamma_{\tilde t_R} = 4 Y_t^2 
+\frac{16}{3} G^2 -8 g_3^2  ~,\nonumber \\
	&&(4\pi)^2 \gamma_{q_L} = h_t^2 + Y_t^2 ~,\nonumber \\
	&&(4\pi)^2 \gamma_{t_R} = 2 h_t^2 + \frac{8}{3} G^2 ~,\nonumber \\
	&&(4\pi)^2 \gamma    _H = 6 h_t^2 ~,  \\
	&& (4\pi)^2 \gamma    _{\tilde H_u} = 3 Y_t^2 ~,\nonumber \\ 
	&&(4\pi)^2 \gamma_{\tilde g} = G^2 ~. \nonumber \\
        &&\gamma_{\tilde W}=\gamma_{\tilde B}=\gamma_{\tilde H_d}=0 \nonumber
\eea
and 
\bea
\label{betaV1}
        &&\beta_{g_u}^{(v)} =
	  \beta_{g'_u}^{(v)}=\beta_{g_d}^{(v)}=\beta_{g'_d}^{(v)} =
	  \beta_{J}^{(v)}=  \beta_{Y_t}^{(v)} = 0  ~,\nonumber \\  
	  &&(4\pi)^2 \beta_{G}^{(v)} = -9~~ g_3^2 G ~, \nonumber\\
          &&(4\pi)^2 \beta_{h_t}^{(v)} = - 8 ~ h_t g_3^2 ~~.
\eea

On the other hand to compute $\beta_{\lambda}^{(v)}$,
$\beta_{Q}^{(v)}$ and $\beta_{K}^{(v)} $ we have found it very
convenient to use the effective potential method~\cite{Gato:1984ya}.
In order to do that we introduce background fields $\phi_c$ and $s_c$
for $H$ and ${\tilde t_R}$ defined as
\begin{equation}
	H \rightarrow 1/\sqrt{2}\left(\begin{array}{c}
			\phi_2 +i \phi_3  \\
			h+\phi_c + i \phi_1  
		 \end{array}\right)          
\end{equation}
\begin{equation}
	{\tilde t_R}^{(\omega)} \rightarrow 1/\sqrt{2}
		\left({\tilde t_{1R}^{(\omega)}}+  \delta^{j3} s_c +i
			{\tilde t_{2R}^{(\omega)}}   \right)  ~, 
\end{equation}
where $\omega$ is the color index.  In this background some fields
acquire a mass and, in particular, the bosonic mass spectrum becomes
\begin{equation}
	\label{bosonic}  
	\begin{array}{rccrl}
	g^{a} :  &  m^2 = 0   &&&	 a=1,2,3   \vspace{1.8mm}\\
	g^{a} :  &  m^2 = s_c^2 ~g_3^2/4  &&&	 a=4,5,6,7   \vspace{1.8mm}\\
	g^{a} :  &  m^2 = s_c^2  ~g_3^2/3  &&&	 a=8  \vspace{1.8mm}\\
	\phi_\omega : &   m^2 = \phi_c^2 \lambda/2 + s_c^2 Q/2  &&& 
\omega=1,2,3  \vspace{1.8mm}\\
	{\tilde t_{1R}^{(\omega)}},{\tilde t_{2R}^{(\alpha)}} : &   
			m^2 = \phi_c^2 Q/2 + s_c^2 K/6  &&& 
\omega=1,2,3 ~ \alpha=1,2  \vspace{1.8mm}\\
		\left(\begin{array}{c}
			{\tilde t_{1R}^{(3)}} \\
			h		
		\end{array}\right) : &
			m^2 = \frac{1}{2}
			\left(\begin{array}{cc}
			s_c^2 K + \phi_c^2 Q   &   2 s_c \phi_c Q\\
			2 s_c \phi_c  Q         &   3 \phi_c^2  
\lambda +s_c^2 Q\\	
		\end{array}\right) 		
	\end{array}
\end{equation}
where we have written only the terms which depend on $\phi_c$ and/or
$s_c$. Analogously the fermionic mass spectrum looks like
\begin{equation}
	\left(\begin{array}{ll}
	b_L^{(3)}   \\
	{\tilde H_u^+} 
\end{array}\right)   ~:~m =
	\left(
	\begin{array}{cc}
 0 & {Y'} \\
{Y'} & 0 \end{array}
	\right)
\end{equation}
\begin{equation}
	\left(\begin{array}{llll}
	t_L^{(1)} \\
	t_R^{(1)\dag} \\
	g^{(4)} \\
	g^{(5)} \\
\end{array}\right)   ~:~m =
	\left(
	\begin{array}{cccc}
0 & h' & 0 & 0 \\
h' & 0 & G' & -i G' \\
0 & G' & 0 & 0  \\
0 & -i G' & 0 & 0 
	\end{array}
	\right)
\end{equation}
\begin{equation}
	\left(	\begin{array}{ll}
	t_L^{(2)} \\
	t_R^{(2)\dag} \\
	g^{(6)} \\
	g^{(7)} 
	\end{array}\right)
	   ~:~m =
	\left(\begin{array}{cccc}
	0 & h' & 0 & 0  \\
	h' & 0 & G' & -i G'\\
	0 & G' & 0 & 0  \\
	0 & -i G' & 0 & 0  
	\end{array}\right)
\end{equation}
\begin{equation}
	\left(\begin{array}{llll}
	t_L^{(3)} \\
	t_R^{(3)\dag} \\
	{\tilde H_u^0}  \\
	g^{(8)} 
\end{array}\right)   ~:~m =
	\left(
	\begin{array}{cccc}
0 & h' & {Y'} & 0 \\
h'& 0 & 0 & -\frac{2 G'}{\sqrt{3}} \\
{Y'} & 0 & 0 & 0 \\
0 & -\frac{2 G'}{\sqrt{3}} & 0 & 0  
	\end{array}
	\right)
	\label{fermionic}
\end{equation}
where $h'=\phi_c h_t / \sqrt{2}$ , $Y'=s_c Y_t / \sqrt{2}$ and $G'=s_c
G /2$.

Using the property of invariance of the effective potential with
respect to the renormalization scale we can write
\bea
\label{invScale}
&&	\tau \frac{dV(\phi_c,s_c)}{d\tau}= 
\nonumber\\
&&\tau 
\frac{d}{d\tau} \left( V_0(\phi_c,s_c) 
	+ \frac{1}{64 \pi^2} \mathrm{STr} \left[\mathcal M^4(\phi_c,s_c)\
	\ln\left[\frac{\mathcal M^4(\phi_c,s_c)}{\tau^2}\right]   \right]  
	+\cdots \right)
        \vspace{2mm}\\
	&&=  \cdots +\frac{1}{8} \beta_\lambda^{(v)} \phi_c^4 
        +\frac{1}{24} \beta_K^{(v)} s^4 
	+\frac{1}{4} \beta_Q^{(v)} \phi_c^2 s_c^2 -  
	\frac{1}{32 \pi} \mathrm{STr} \left[\mathcal M^4(\phi_c,s_c)\right]=0 
	\nonumber ~,
\eea
where the ellipses stand for terms we are not interested in,
$V_0(\phi_c,s_c)$ is the tree--level scalar potential of
(\ref{lagreff}) in the presence of the background fields $\phi_c$ and
$s_c$, $\mathcal M^2(\phi_c,s_c)$ is the mass spectrum of the fields
written in (\ref{bosonic})-(\ref{fermionic}) and, finally, for a given
function $f(\mathcal M^2)$ of the squared mass matrix of all fields in
the theory, $\mathrm{STr} f(\mathcal M^2)
\equiv\mathrm{Tr}\sum_J(-1)^{2J}(2J +1) f(M_J^2)$.  Therefore in
(\ref{invScale}) $\beta^{(v)}_\lambda$, $\beta^{(v)}_Q$ and
$\beta^{(v)}_K$ are put easily in evidence by the expansion in powers
of $s_c$ and $\phi_c$ of $\mathrm{STr}\mathcal M^4(\phi_c,s_c)$.
Furthermore for our purposes only the terms $M_J^4(\phi_c,s_c)$
proportional to $s_c^4$, $\phi_c^4$ or $s_c^2 \phi_c^2$ are
interesting and consequently we can ignore in $M_J^2(\phi_c,s_c)$ the
dependence on dimensional couplings, as we have done in
(\ref{bosonic})-(\ref{fermionic}). After performing the corresponding
expansions we get the $\beta$-functions
\bea
\label{betaV2}
	&&(4\pi)^2 \beta_\lambda^{(v)} = 12 \lambda^2 +6 Q^2 -12 h_t^4 ~,
\nonumber \\
	&&(4\pi)^2 \beta_Q^{(v)} = K Q +3 \lambda Q+ 4 Q^2 
		- \frac{32}{3} G^2 h_t^2 - 4 Y^2 h_t^2 ~, \nonumber \\
	&&(4\pi)^2 \beta_K^{(v)} = 12 Q^2+\frac{14}{3}K^2 +13 g_3^4
		- \frac{88}{3}G^4 -24 Y_t^4   ~.
\eea
Finally by plugging the results (\ref{betaF}), (\ref{betaV1}) and
(\ref{betaV2}) into (\ref{beta}), we find the result which was
anticipated in (\ref{stopbeta}).

In order to complete our renormalization picture we will compute now
the running of the masses. By using standard diagrammatic methods we
obtain
\bea
\label{betaV3}
        &&\beta_{M_1}^{(v)} = \beta_{M_2}^{(v)} = \beta_{\mu}^{(v)}= 0
        ~,\nonumber \\
        &&(4\pi)^2 \beta_{M_3}^{(v)} = -18 g_3^2 M_3 ~, \\
        &&(4\pi)^2 \beta_{m^2}^{(v)} = -6 Q~m_U^2  ~, \nonumber \\
        &&(4\pi)^2 \beta_{M_U^2}^{(v)} = -\frac{32}{3} M_3^2 G^2 
         +\frac{8}{3}K M_U^2 -4 m^2 Q -4 Y_t^2 ~\mu^2 \nonumber ~,
\eea
and using (\ref{beta}) and (\ref{betaF}) we find the expressions
in Eq.~(\ref{massbeta}).

\end{document}